\documentclass[12pt,letterpaper]{article}

\usepackage[margin=1in]{geometry}
\usepackage{setspace}
\usepackage{lmodern}
\usepackage[T1]{fontenc}
\usepackage{microtype}
\usepackage{amsmath, amssymb, amsthm}
\usepackage{graphicx}
\usepackage{booktabs}
\usepackage{threeparttable}
\usepackage{array}
\usepackage{caption}
\usepackage{subcaption}
\usepackage{natbib}
\usepackage[colorlinks=true,allcolors=blue]{hyperref}
\usepackage{xcolor}
\usepackage[section]{placeins}   % keep floats within their sections
\usepackage{fancyhdr}

\usepackage{etoolbox}
\AtBeginEnvironment{tabular}{\singlespacing}
\hypersetup{pdfauthor={}, pdftitle={}}

\newcommand{\R}{\textsf{R}}
\newcommand{\Stata}{Stata}
\newcommand{\pkg}[1]{\textsf{#1}}
\newcommand{\code}[1]{\texttt{#1}}

\newif\ifanon
\anonfalse

\ifanon
\else
\fi

\fancypagestyle{plain}{\fancyhf{}%
  \fancyfoot[L]{\small\scshape Binned Test Scores}%
  \fancyfoot[R]{\small\thepage}}

\title{Faster Estimates from Binned Test Scores---With Discussion of Normality, Bias, and Reliability}
\ifanon
  \author{}
\else
  \author{%
    Paul T.~von Hippel\thanks{Email: \href{mailto:paulvonhippel@utexas.edu}{paulvonhippel@utexas.edu}.} \\
    {\small LBJ School of Public Affairs, University of Texas at Austin}
  }
\fi
\date{}

\begin{document}

{\singlespacing\maketitle}
\vspace{-2.5em}

%% Abstract--------------------------------------------------------
%% Single-spaced, as journals typeset abstracts; this also keeps the
%% title, byline, abstract and keywords together on page 1.
\begin{abstract}
\singlespacing
\noindent
Education agencies often summarize test score distributions by counting how many students scored in 3 to 5 \emph{bins}. Means and standard deviations (SDs) can be estimated from bin counts using a HETOP model, which assumes that scores are normally distributed within each school or district. However, most HETOP software takes minutes to hours if the number of schools or districts is large. 

We present a \emph{fast HETOP} algorithm that produces HETOP estimates in less than a millisecond per district.\ifanon\else{} The algorithm is implemented in the \href{https://cran.r-project.org/web/packages/binest/refman/binest.html\#topic+fast\_hetop}{\code{fast\_hetop}} function of the \pkg{binest} package for \R{}.\fi

HETOP estimates of the mean are highly reliable; estimates of the SD are not as good. Maximum likelihood estimates are generally excellent but can be biased and unreliable if scores are not normal or scores are concentrated in the top or bottom bin. Empirical Bayes shrinkage improves unreliable estimates. Standard errors are smaller if the data represents the whole population of students rather than a sample. 

HETOP estimates are usable for many purposes, but it would be better if education agencies reported means and SDs directly.

\end{abstract}


%% Named colleagues and the author's own role are identifying, so this
%% block is suppressed under \anontrue.  JREE asks that statements of
%% credit go on the separate cover page for the anonymous submission.
\ifanon\else
\textbf{Acknowledgments and AI disclosure:} The author is responsible for all questions, ideas, writing, and organization. The author thanks David Hunter and Benjamin Shear for insightful conversations and suggestions. Claude helped the author derive formulas, develop algorithms, write code, and find and review old statistics papers. The author improved the intelligibility of mathematical derivations, checked code and simulation results, and verified that fast HETOP estimates matched HETOP estimates from older software.
\fi

%% The cover page carries no running head or page number, and the
%% Introduction starts on page 1.
\thispagestyle{empty}
\clearpage
\setcounter{page}{1}

%% Body------------------------------------------------------------
\section{Introduction}

Education agencies often categorize test scores into 3 to 5 different score ranges, or \emph{bins}, delimited at certain score thresholds, or \emph{cuts}.\footnote{We favor the words \emph{bins} and \emph{cuts} because they are short, clear, and distinctive. Confusingly, the classical statistics literature uses the word \emph{group} for what we call bins \citep[e.g.,][]{heitjan1989,hald2001,haitovsky1982}, while recent education papers use the same word \emph{group} for what we call districts \citep[e.g.,][]{reardon2017,shear2019,lockwood2018}. Bins have also been called \emph{brackets}, \emph{ranges}, \emph{intervals}, or \emph{coarsened} or \emph{ordinal} categories. Cuts have also been called \emph{thresholds}, \emph{boundaries}, \emph{borders}, and \emph{edges}.} In 2017--18, for example, the Texas Education Agency grouped 6th grade math scores into 4 bins: the lowest bin, labeled \emph{unsatisfactory} (below the first cut of 1536), and higher bins indicating that students \emph{approached}, \emph{met}, or \emph{mastered} grade-level standards. 

\begin{figure}[!htbp]
\centering
\includegraphics[width=\textwidth]{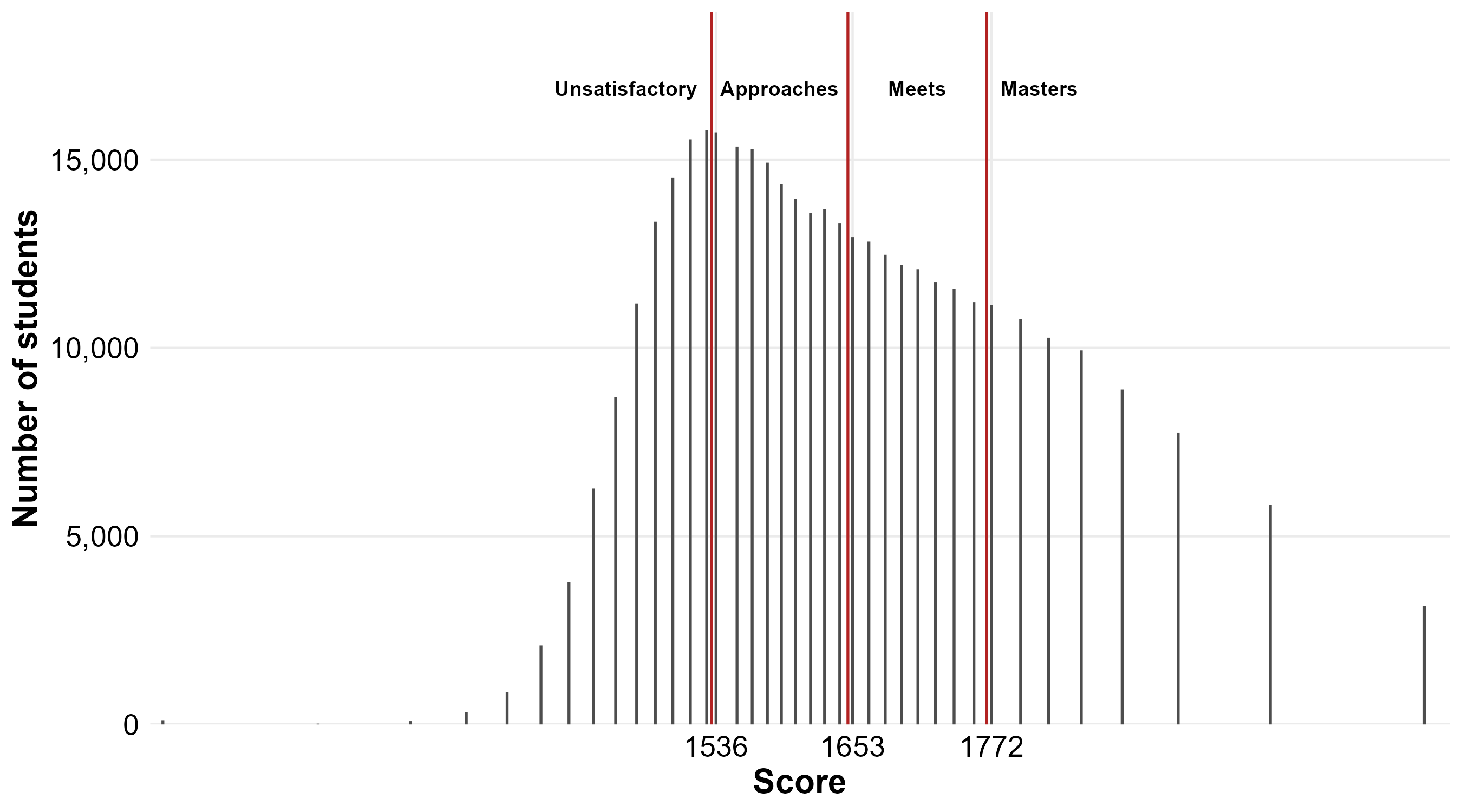}
\caption{Statewide distribution of 6th grade math scores on the State of Texas Assessments of Academic Readiness (STAAR) in 2017--18. The red lines show the cut scores between bins.}
\label{fig:state-dist}
\end{figure}

In addition to the statewide score distribution (Figure~\ref{fig:state-dist}), the Agency published a summary table counting how many students scored in each bin (Table~\ref{tab:tx-bins}). The Agency published similar binned counts for the whole state, for each region, district, and school, for all students and for 10 student subgroups: African American, Hispanic, White, American Indian, Asian, Pacific Islander, Multiracial, Special Education, Economically Disadvantaged, and English Language Learners \citep{tea2018}.

\begin{table}[!htbp]
\centering \small
\caption{Binned summary of the Texas score distribution shown in Figure~\ref{fig:state-dist}.}
\label{tab:tx-bins}
\begin{tabular}{lrrr}
\toprule
Proficiency category&Score range                & Students & Percentage \\
\midrule
Unsatisfactory & $<$1536 & 92{,}627  & 23.9\% \\
Approaches &1536--1652 & 130{,}187 & 33.6\% \\
Meets &1653--1771 & 97{,}065  & 25.0\% \\
Masters & $\ge$1772 & 67{,}753  & 17.5\% \\
\midrule
Total                     &            & 387{,}632 & 100.0\% \\
\bottomrule
\end{tabular}
\end{table}
\FloatBarrier

While binned summaries give an impression of how well students performed relative to state standards, reliance on binned summaries makes it challenging to estimate effects or trends. The location of cuts is arbitrary, and the relationship between a shift in the score distribution and a change in bin percentages is not straightforward: it depends on the score distribution, on the position of the cuts, and on how many students scored near each cut \citep{ho2008,holland2002}. In Texas, for example, there are more students near the lowest cut than near the highest cut, so an upward shift of $0.2$ standard deviations (SD) would reduce students in the ``unsatisfactory'' bin by 8 percentage points, while increasing students in the ``masters'' bin by only 3 percentage points. This might create the impression that low-scoring students had improved more than high-scoring students, even if all students' scores improved by exactly the same amount.

\subsection{Estimation with binned test scores}
Researchers working with binned test scores often want statistics such as the mean and SD for each district, school, or subgroup. Texas reports mean test scores for each district, but 36 US states do not, and hardly any states report SDs for schools or districts. So researchers must often estimate means and SDs from binned counts.

Estimation from binned data typically requires a simplifying assumption about the shape of the underlying distribution, and the fewer bins there are, the simpler the assumed distribution must be. In general, the number of parameters in the assumed distribution must be less than the number of bins \citep{kulldorff1961}. 

The normal distribution, which has just 2 parameters (mean and SD), is a natural choice for test score data reported using just 3 to 5 bins. The normal distribution is often considered a serviceable approximation for test scores, although it is only an approximation and can be far from a perfect fit to some test score data \citep{ho2015,holland2002}. The score distribution in Texas (Figure~\ref{fig:state-dist}), for example, is visibly non-normal, with positive skewness ($0.76$) and values that are discrete rather than continuous.\footnote{Discreteness comes from the fact that the test had 38 items, and the reported Rasch-scaled score is a one-to-one function of the number of items answered correctly \citep{tea2018}.}

\subsection{Fast estimation of the normal (HETOP) model}

A popular model for estimating means and SDs from binned data assumes that scores are normally distributed within each district (or school or subgroup), that each district has its own mean and SD, and that all districts share the same cuts and bins. In education, this is called the \emph{heteroskedastic ordered probit (HETOP)} model \citep{reardon2017,lockwood2018}. It is \emph{ordered probit} because, within each district, the data can be modeled by a probit regression of the ordered bin categories on the underlying normal test scores. It is \emph{heteroskedastic} because each district has a different SD.

Speed in fitting the HETOP model is essential, because education researchers often want estimates not just from a single binned distribution but from hundreds or thousands of binned distributions---for example, every subgroup within every school or district in a large state. Unfortunately, preexisting HETOP software slows down dramatically as the number of distributions being estimated grows large. When fit to over 1,000 districts in Texas, for example, we will shortly see that the fastest HETOP software takes several minutes, and the slowest may fail to finish at all. If we ran not just every district in Texas but all 7,000 elementary schools, separately for 3 tested grade levels and 10 student subgroups---making over 200,000 binned distributions in all---even the fastest existing software might choke. Perhaps in part for this reason, the Stanford Education Data Archive, the leading source of HETOP estimates at the district level, has chosen not to distribute HETOP estimates for individual schools \citep{fahle2021}.

In this article, we present a \emph{fast HETOP} algorithm that produces practically identical HETOP estimates for the same 1,000+ Texas districts in less than a second. The algorithm is not entirely new, but builds on efficient formulas and algorithms developed for binned data from the 1920s to the early 1980s \citep[e.g., ][]{fisher1922,stevens1948,gjeddebaek1949,gjeddebaek1956,kulldorff1958,kulldorff1958b,schader1984}, when efficiency was at a premium because computers, when they were accessible at all, were orders of magnitude slower, and many calculations were still carried out by hand.\ifanon\else{} We implement the algorithm in the \href{https://cran.r-project.org/web/packages/binest/refman/binest.html\#topic+fast\_hetop}{\code{fast\_hetop}} function of the \pkg{binest} package for \R{} \citep{vonhippel2026}.\fi

Fast HETOP makes it practical to fit thousands of districts, schools, or subgroups. Speed makes it faster to test HETOP software and add new features. And speed makes it practical to combine HETOP with time-consuming procedures that require estimating each district many times---such as the bootstrap or multiple imputation. 

Speed also makes it easier to explore the properties of HETOP estimates in simulated and empirical data. We evaluate the quality of HETOP estimates against district data for a test given in Texas. Like past investigators, we find that HETOP estimates are very strongly correlated with true means, but somewhat less correlated with true SDs \citep{reardon2017,shear2021,fahle2021}. We also find that HETOP estimates can be biased and unreliable when scores are not normal, or in low-information districts where scores are concentrated in the top or bottom bin \citep[cf. ][]{kulldorff1958,gjeddebaek1959b,reardon2017,lockwood2018}. Estimates from low-information districts can be improved by using Bayesian methods that shrink unreliable estimates toward more reliable estimates from other districts.

The ordering of this article prioritizes the interests of applied researchers. We start by analyzing empirical test score data from Texas---addressing practical issues such as the speed of software, the accuracy of estimates, and the sensitivity of estimates to non-normality and different analytic choices. We then revisit those questions in districts simulated to meet HETOP's assumption of normally distributed test scores. Finally, we explain what makes fast HETOP fast---keeping our explanation at a high, intuitive, accessible level, and consigning detailed formulas and algorithms to the Appendices.

\section{Evaluation in Texas}

We start by fitting the HETOP model to the Texas math score data from Figure~\ref{fig:state-dist}. The Texas Education Agency published the count of scores in each of the four proficiency bins for every one of the state's $1{,}151$ school districts \citep{tea2018}. Each district tested between 5 and $13{,}191$ 6th grade students, with an average of 337 6th grade students per district. 

In addition to bin counts, the Agency also published the mean ($1640$) and SD ($139$) of scores across all $N = 387{,}632$ sixth graders, and the mean (but not the SD) of scores for each district. The published district means represent true means against which the accuracy of HETOP-estimated means can be evaluated.

Of the $1{,}151$ districts, $17$ small districts had nonzero counts in only 1 or 2 bins, fewer than the 3 necessary to identify 2 parameters---the district mean and SD---under the HETOP model. We will limit our results to the $1,134$ identified districts. Appendix~\ref{app:unident} describes approximations that can produce estimates for unidentified districts.

\subsection{Runtime} \label{sec:runtime}

Table~\ref{tab:software} shows the time required to run $1,134$ Texas districts through 5 different HETOP programs. We will discuss the programs' similarities and differences later; for now, the important contrast is between the fast HETOP programs, which estimate one district at a time, and older \emph{joint HETOP} programs, which estimate all $2,268$ means and SDs together---that is, jointly.

Fast HETOP runs approximately 1,000 times faster than any joint HETOP implementation. Fast HETOP ran all $1{,}134$ districts in 0.2 to 0.3 seconds, vs. nearly 4 minutes for the fastest joint HETOP program, and 27 minutes for the next-fastest. The slowest joint HETOP program did not finish and was stopped after an hour.

\begin{table}[!ht]
	\centering \small
	\begin{threeparttable}
	\caption{Five software implementations of the HETOP model}
	\label{tab:software}
	\setlength{\tabcolsep}{2pt}
	\footnotesize
	\begin{tabular}{llrrllll}
		\toprule
		Algorithm & Estimates & \shortstack[r]{Runtime\\(sec)} & Corr. & Function & Package & Software & Programmer \\
		\midrule
		%% The function, package and author identify the authors, so the
		%% anonymous version withholds them.
		\ifanon
		Fast  & ML & $0.3$      & $0.97$ & --- & --- & \R{}     & this article \\
		\else
		Fast  & ML & $0.3$      & $0.97$ & \code{fast\_hetop} & \pkg{binest} & \R{}     & \citet{vonhippel2026} \\
		\fi
		Joint & ML & $223$      & $0.97$ & \code{hetop}       & \pkg{hetop}  & \Stata{} & \citet{shear2019} \\
		Joint & ML & $>3{,}600$ & did not finish    & \code{mle\_hetop}  & \pkg{HETOP}  & \R{}     & \citet{lockwood2019} \\
		\midrule
		\ifanon
		Fast  & Bayes & $0.3$     & $0.98$ & --- & --- & \R{}  & this article \\
		\else
		Fast  & Bayes & $0.3$     & $0.98$ & \code{fast\_hetop} & \pkg{binest} & \R{}  & \citet{vonhippel2026} \\
		\fi
		Joint & Bayes & $1{,}620$ & $0.98$ & \code{fh\_hetop}   & \pkg{HETOP}  & \R{}, JAGS  & \citet{lockwood2019} \\
		\bottomrule
	\end{tabular}
	\begin{tablenotes}[flushleft]\footnotesize
		\item \textit{Note.} ML is maximum likelihood. Correlation is measured against true district means.
	\end{tablenotes}
	\end{threeparttable}
\end{table}

\begin{figure}[!ht]
\centering
\includegraphics[width=0.9\textwidth]{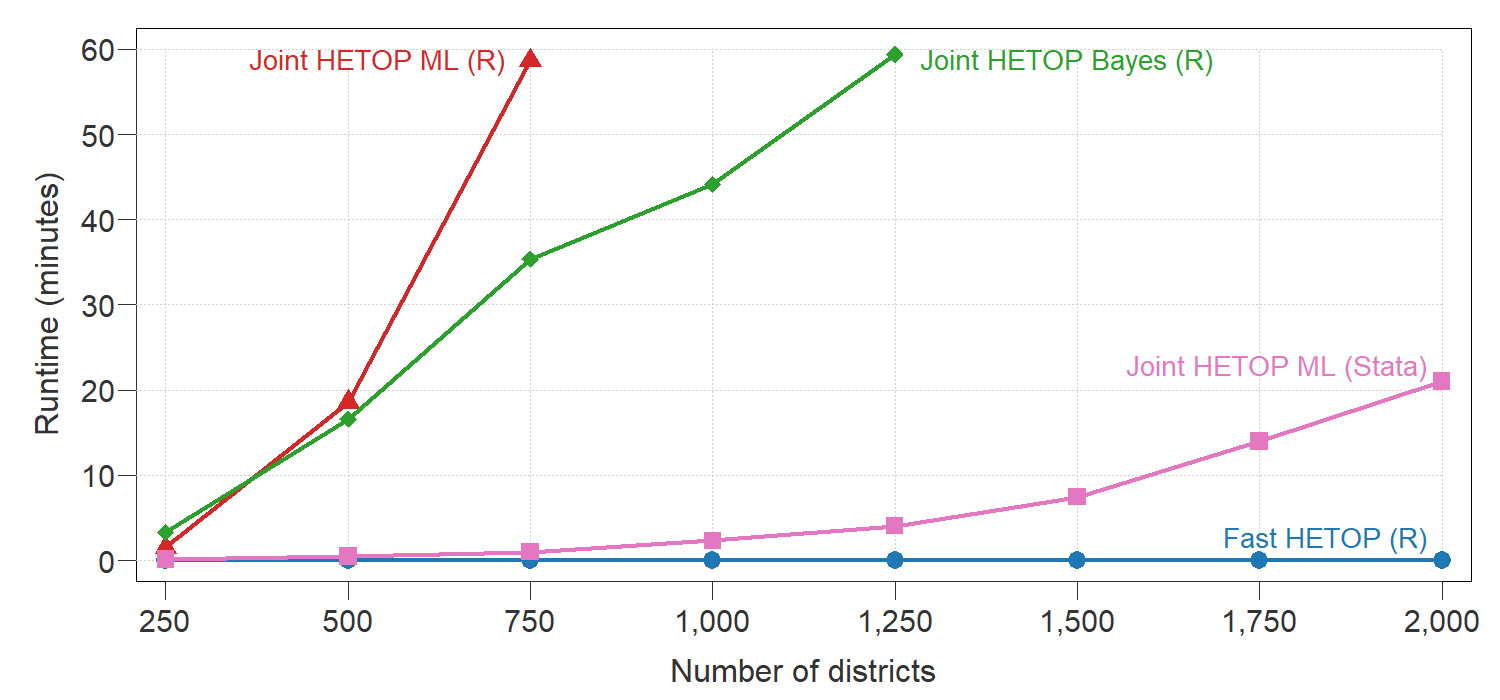}
\caption{Runtime versus the number of districts for four different HETOP programs. Runtimes measured on a Dell Latitude laptop with an Intel Core Ultra 7 165U processor (2.10\,GHz) and 32\,GB of RAM, running Windows 11, R 4.5.1, Stata/MP 18.0, and JAGS 4.3.2.}
\label{fig:scaling}
\end{figure}

Figure~\ref{fig:scaling} shows how each program's runtime scales with the number of districts. We drew samples of $250$ to $2{,}000$ districts, with larger samples made possible by sampling with replacement. Fast HETOP ran the largest sample of $2,000$ districts in 0.6 seconds. Runtime increased proportionately to the number of districts, implying that fast HETOP would take just a minute to run $200,000$ distributions---for example, every student subgroup in three grade levels in every Texas elementary school.

Every joint HETOP implementation ran much more slowly, and runtime grew more quickly than the number of districts. At $250$ districts, all joint HETOP programs ran in 3 minutes or less, but doubling the number of districts increased some programs' runtimes by a factor of 5 or 10. One joint HETOP program reached an hour of runtime at $750$ districts, and another reached an hour at $1,250$ districts. The fastest joint HETOP program ran $1,000$ districts in about 2 minutes, but $2,000$ districts took about 20 minutes; that is, doubling the number of districts increased runtime more than ten times. 

\FloatBarrier

\subsection{Maximum likelihood point estimates} \label{sec:empirical}
\paragraph{Equivalence of fast and joint ML estimates.}
Both fast and joint HETOP can estimate means and SDs using maximum likelihood (ML). The joint and fast HETOP estimates are theoretically identical because they maximize the same likelihood. Numerically, estimates can differ microscopically because both fast and joint HETOP iterate toward the maximum of the likelihood without quite reaching it. In the Texas data, using the default thresholds, the correlation between fast HETOP and joint HETOP estimates was over $0.999999$, and we could make fast and joint HETOP estimates agree as closely as desired by tightening the threshold for convergence.

\begin{figure}[!ht]
	\centering
	\includegraphics[width=\textwidth]{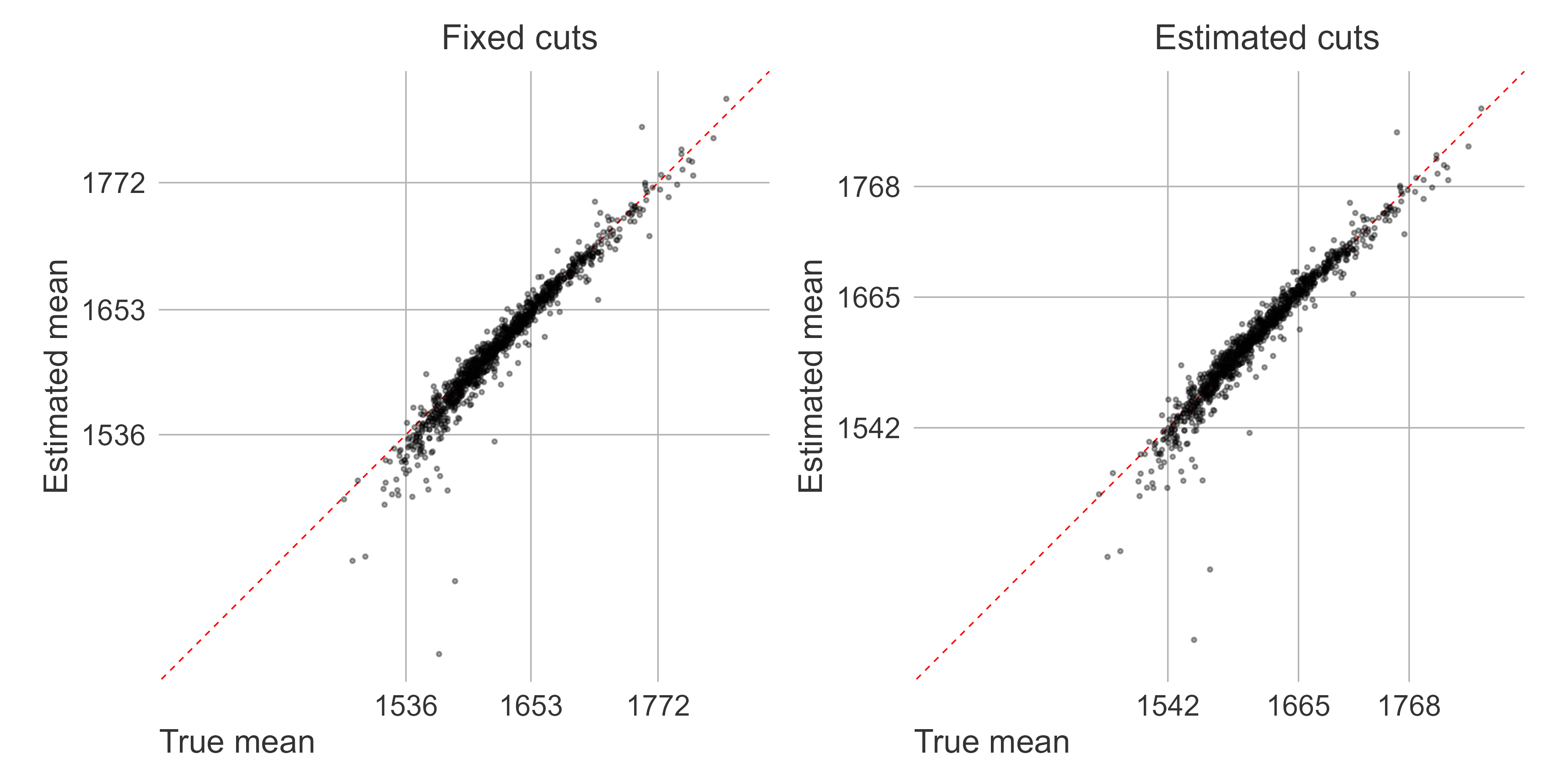}
	\caption{Fast HETOP estimates of district means ($Y$) against true district means ($X$). Perfectly accurate estimates would lie on the dashed red diagonal ($Y = X$). The left panel fits the HETOP model around fixed cut scores (indicated by grid lines). The right panel fits the model to the state mean and SD, and estimates the cut scores.}
	\label{fig:tx-scatter-both}
\end{figure}

\paragraph{Accuracy of ML estimates}
Figure~\ref{fig:tx-scatter-both} plots true district means (X) against ML estimates of district means produced by fast HETOP. Overall, the mean estimates are excellent; the correlation between true and estimated means is very strong ($0.97$).

\FloatBarrier

\subsubsection{Comparison of estimates using fixed vs. estimated cuts}

To identify district means and SDs, the HETOP model needs additional information to fix each district's position in the state distribution. We can either 
\begin{enumerate}
	\item \emph{Fixed cuts.} fix the cut scores at their published values, or
	\item \emph{Estimated cuts.} fix the mean and SD for the state, and let the model estimate the cut scores.
\end{enumerate}

When fixing the state mean and SD, it is common to choose a mean of 0 and an SD of 1, so that estimates can be interpreted on a standardized scale. In our Texas example, though, we will use the published mean and SD for the state, so that the estimated cuts can be compared to the fixed cuts directly on the test's native scale.

For most purposes, it makes little difference whether the HETOP model is identified using fixed or estimated cuts. The scatter cloud of estimated means against true means looks practically identical, including a few outlying points, and mean estimates that use fixed cuts are almost perfectly correlated ($0.9998$) with mean estimates that use estimated cuts.

There is one material difference, though. Mean estimates are 6 points (0.04 SD) higher, on average, when we use estimated rather than fixed cuts. This is because fixed-cut approaches must use the provided cuts and can under- or overestimate the mean, while the estimated-cut approach puts the cuts wherever it needs to, in order to match the statewide mean and SD. In the Texas data, the lowest two cuts are 6 to 12 points ($0.04$ to $0.09$ SD) above the corresponding fixed cuts. The highest estimated cut ($1768$) is below the highest fixed cut ($1772$), but only by $4$ points ($0.03$ SD).

So in Texas, fixing the cuts yields a slight downward bias, which using estimated cuts eliminates---at least on average.

\subsubsection{Conditional bias of estimates} \label{sec:tx-cond-bias}

\begin{figure}[!ht]
	\centering
	\includegraphics[width=\textwidth]{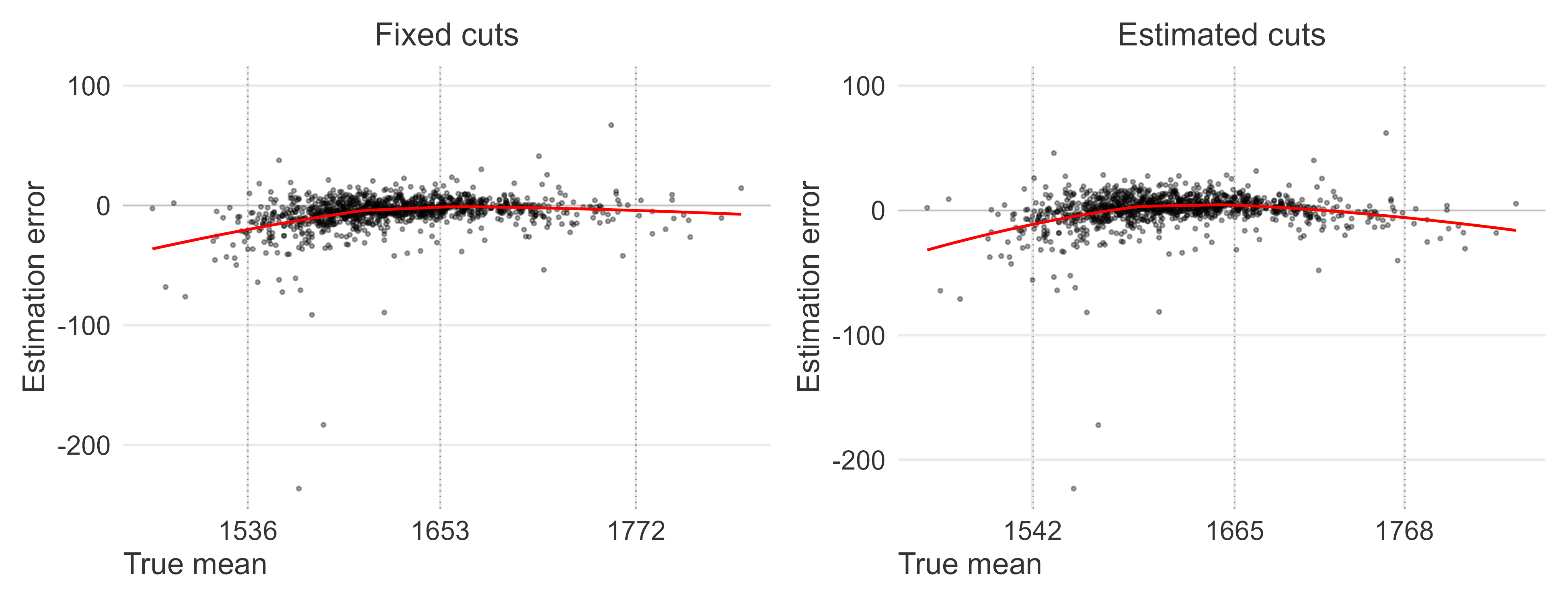}
	\caption{Estimation error ($Y$) against true district means ($X$), fitted with a red loess curve. The bias (average error) is near zero through the middle of the distribution, but negative at both extremes.}
	\label{fig:tx-err-vs-truth}
\end{figure}

Although the fixed-cut estimates have about the right mean on average, they are biased in the extremes of the distribution. Figure~\ref{fig:tx-err-vs-truth} displays the bias by plotting each district's estimation error against the true district mean. With cuts fixed, estimation error in the lowest-scoring decile of Texas districts is $-16$ points ($-0.12$ SD), on average, and the worst estimate is $236$ points below its district's true mean. With estimated cuts, the pattern of bias is similar, though the negative estimation errors are slightly smaller because the estimated means are slightly higher.

\FloatBarrier

\subsubsection{Non-normality as a source of bias}

\begin{figure}[!ht]
	\centering
	\includegraphics[width=\textwidth]{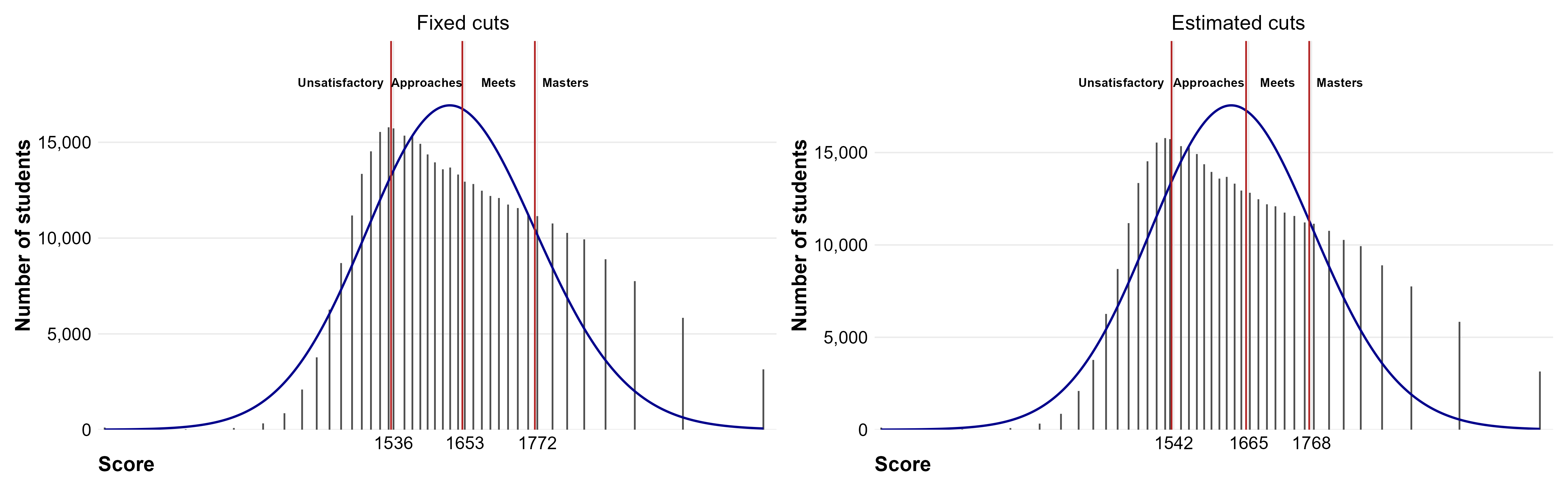}
	\caption{HETOP's fitted mixture of district normals, drawn over the empirical Texas score distribution. Left: the model fitted with fixed cuts. Right: the model with estimated cuts.}
	\label{fig:tx-mixture}
\end{figure}

Where does this bias come from? Non-normality of test scores is one source. Figure~\ref{fig:tx-mixture} compares the empirical test score distribution to the distribution fitted by the HETOP model. The fitted distribution looks normal, but it is actually a student-weighted \emph{mixture} of normal distributions fitted to each district.\footnote{In general, the fitted normal mixture distribution will be dominated by normal distributions fitted to the state's largest districts. Unless those districts differ dramatically in their means or SDs, the normal mixture will itself be nearly normal.}  Unlike the empirical distribution, which is positively skewed (+0.76), the fitted normal mixture distribution is practically symmetric (with a skewness of just 0.08 with fixed cuts, or 0.04 with estimated cuts).

Because HETOP estimates are drawn from the fitted distribution rather than the empirical distribution, discrepancies between the fitted and empirical distribution can cause bias:
\begin{itemize}
	\item The fitted distribution has a longer left tail than the empirical distribution. So estimates for the lowest-scoring districts, where scores concentrate in the left tail, tend to be too low.
	\item The fitted distribution also has a lighter right tail than the empirical distribution. So estimates for the highest-scoring districts, where scores concentrate in the right tail, tend to be too low as well.
\end{itemize}

The HETOP model does not assume that the state distribution is normal, but only assumes that the distribution is normal within each district. We can check this assumption with a chi-square statistic (Appendix~\ref{app:gof}), which tests the null hypothesis that a district's counts could come from binning a normal distribution. This test rejects normality for 19\% of Texas districts, and it is likely that more districts are non-normal but too small for the test to have adequate power. So even without Figure~\ref{fig:tx-err-vs-truth}, we would question the model's assumption of within-district normality.

\subsection{Can the score distributions be transformed to normality?} \label{sec:misspec}

What is attractive about HETOP mean estimates is their strong correlation with true means published by education agencies and vendors (Figure~\ref{fig:tx-scatter-both}; \citealp{kuhfeld2019,fahle2021,shear2021,reardon2017}). Yet the HETOP model assumes normality, while test scores are not normal in Texas or other states \citep{ho2015,holland2002}. 

Presentations of the HETOP model often address non-normality by arguing that HETOP estimates are not meant to be interpreted on the published test score scale, but on a scale that has been implicitly transformed to normality within districts. This interpretation, called \emph{respective normality}, assumes that there exists a single monotone transformation $Z_g=t(Y_g)$, that transforms a possibly non-normal test score $Y_g$ into a new score $Z_g$ that is normal within every district $g$ \citep{ho2012,reardon2017}.

Notice that the respective normality argument does not really apply to HETOP estimates obtained using fixed cuts. Any transformation that substantially changes the shape of the distribution would need to respace the cuts---but the cuts cannot be respaced if they are fixed. 

Using estimated cuts opens the door to transformation because estimated cuts can be respaced. Yet in Texas, there was little difference between the fixed and estimated cuts, and the distribution of mean estimates had the same shape whether cuts were estimated or fixed. While we cannot directly observe HETOP's transformation of Texas' test scores, the similarity of fixed- and estimated-cut estimates suggests that any transformation must have been very mild.

Why didn't HETOP use a stronger transformation to normalize Texas' scores? In general, there may not be a transformation that can normalize every district. Test scores in Texas are discrete, not continuous, and no monotone transformation can turn a discrete distribution into a continuous distribution like the normal. And even if scores were continuous, it is unlikely that the \emph{same} transformation could normalize every district. For example, if two districts have a similar score range but opposite skewness---one skewed to the right, one to the left---then the same transformation cannot normalize both of them. Even if two districts have the same shape, the same transformation may not normalize them if they have different score ranges. For example, suppose one district ($g$) has scores $Y_g$ that are log-normal, while another district ($h$) has the same scores shifted half a point to the right, $Y_h=Y_g+0.5$. Then a log transformation will perfectly normalize the log-normal district $g$, removing its skewness, but the same log transformation will leave the shifted district ($h$) with substantial positive skewness ($0.8$).

\subsection{Empirical Bayes estimates} \label{sec:eb}

How might we improve estimates in the tails of the distribution? One option is to replace ML estimates with empirical Bayes (EB) estimates, which shrink estimates toward the grand mean (Appendix~\ref{app:eb}). In general, we expect EB shrinkage to make the most difference to noisy estimates that are far from the grand mean---and these are the districts with the noisiest and most biased estimates in Texas.

\begin{figure}[!ht]
	\centering
	\includegraphics[width=\textwidth]{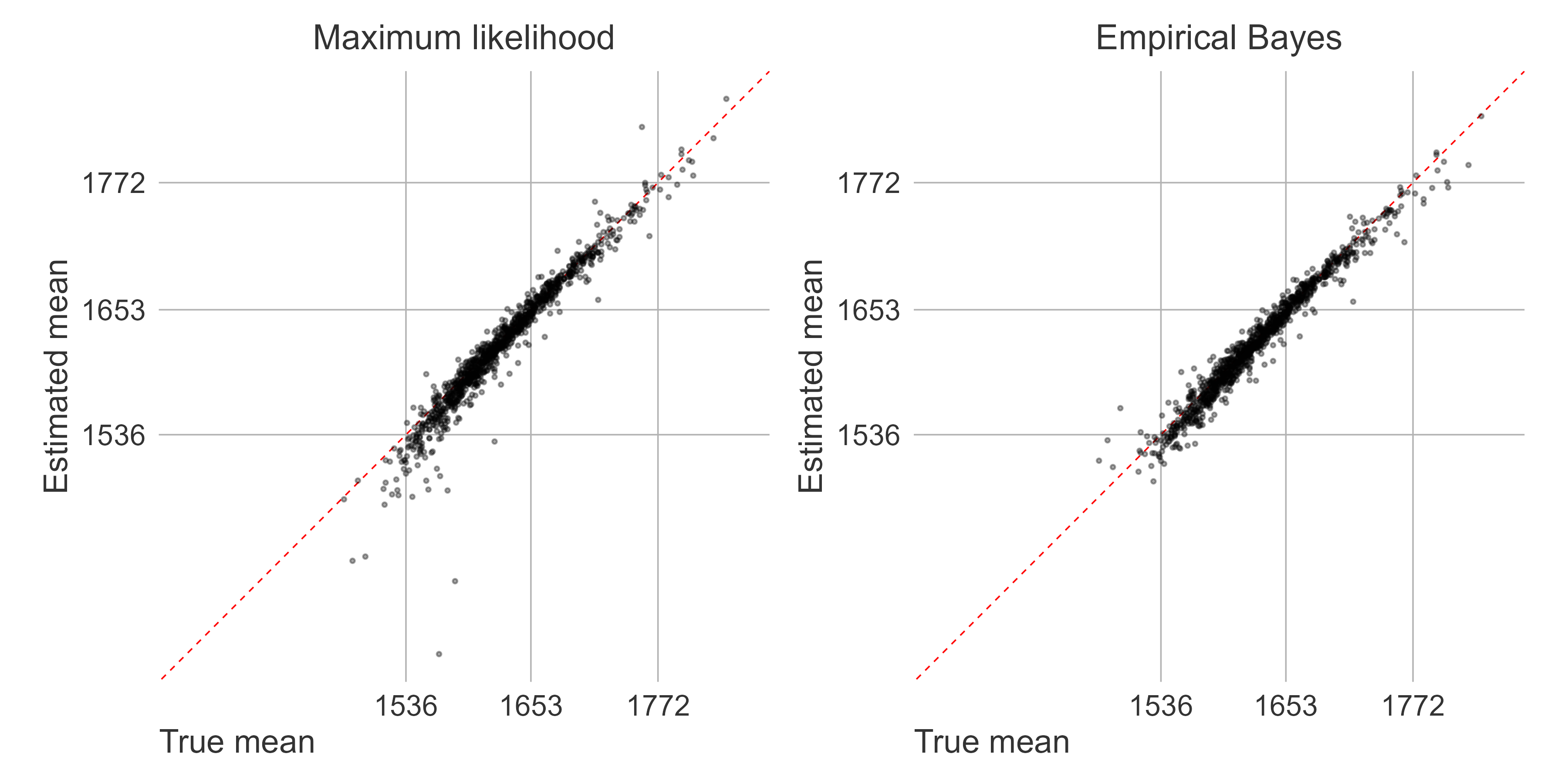}
	\caption{Estimated district means ($Y$) against true district means ($X$), with the cut scores fixed and marked by grid lines. Left: maximum likelihood estimates. Right: shrunken empirical Bayes estimates. Perfectly accurate estimates would lie on the dashed red diagonal.}
	\label{fig:tx-eb-pop}
\end{figure}

Figure~\ref{fig:tx-eb-pop} compares ML estimates (left) to EB estimates (right). The EB estimates are dramatically better in the lower tail, pulling the lowest underestimates up toward the true district means, or slightly higher. In the upper tail, EB estimates are a little worse, pulling most estimates slightly below the true district means, but lower-tail districts get better more than upper-tail districts get worse. In the middle of the distribution, EB makes little difference because the ML estimates are already close to the grand mean that EB shrinks toward. 

Overall, EB reduces the root mean squared error of the mean estimates---from $16$ points for ML estimates to $11$ points for EB estimates, with the worst district estimation error improving from $-236$ to $-55$ points.

\FloatBarrier

\subsection{Standard errors} \label{sec:err-sources}

In addition to any bias, estimates vary randomly around their expected values. There are two sources of random estimation error.

\paragraph{Sampling error}

One source is ordinary \emph{sampling error}, which comes from having only a sample rather than the full population of students. Sampling error exists in federal data like the National Assessment of Educational Progress (NAEP), given to a sample of students in each state. But sampling error is not a concern on state-required tests in Texas and other states, where schools are required to test the whole population (with modifications for students with special needs or limited English).

\paragraph{Binning error}

Another source is \emph{binning error}, which comes from having bin counts rather than individual student scores. Binning error is present even if we have scores for the whole population. In Figure~\ref{fig:tx-scatter-both}, we showed mean scores for the 6th grade population of nearly every district in Texas. The difference between the true means and estimated means was due entirely to binning error.

In Appendix~\ref{app:ml-ci}, we derive formulas to estimate the standard errors (SE) due to binning and sampling. Our fast HETOP software can report a binning SE when we have population data, as in Texas, or a binning+sampling SE when we have sample data. When we shrink point estimates with empirical Bayes, the SEs shrink as well, but not as much (Appendix \ref{app:eb}).

\subsection{Confidence intervals} \label{sec:tx-coverage}

Confidence intervals (CIs) add and subtract 2 SEs from the point estimate of the mean (Appendix~\ref{app:ml-ci}). Table~\ref{tab:tx-coverage} reports CIs for the district means in Texas, using SEs that reflect only binning error because we have a population.

Although these confidence intervals are designed to have $95\%$ coverage, empirically they cover only 
$77$ to $79\%$ of true district means. Coverage remains low whether we use fixed or estimated cuts, and whether we use ML or EB estimates. 

Coverage is low because Texas scores are not normally distributed within districts. If scores were normally distributed, HETOP confidence intervals would have close to $95\%$ coverage, as we will see in Section~\ref{sec:sim-coverage}. But with non-normal data, coverage can be substantially lower. In Texas, non-normality biases the point estimates in some parts of the distribution, and the bias can be large enough to take point estimates more than 2 SEs away from the true mean. The SE estimates are also biased, since their formulas use the biased point estimates (Appendix~\ref{app:ml-ci}).

\begin{table}[!ht]
	\centering
	\caption{Coverage of nominal $95\%$ confidence intervals for district means in Texas.}
	\label{tab:tx-coverage}
	\begin{tabular}{llr}
		\toprule
		Estimator & Cuts & Coverage \\
		\midrule
		Maximum likelihood & fixed     & $78\%$ \\
		Maximum likelihood & estimated & $77\%$ \\
		Empirical Bayes    & fixed     & $78\%$ \\
		Empirical Bayes    & estimated & $79\%$ \\
		\bottomrule
	\end{tabular}
\end{table}
\FloatBarrier

\section{Evaluation in simulated normal data} \label{sec:sim}

This section repeats the evaluation using data simulated to honor HETOP's assumption that scores are normally distributed within districts. The simulation will tell us whether estimates perform better on HETOP's home turf. The simulation will also tell us how well HETOP estimates the SD, which Texas did not report, as well as the mean, which Texas reported as well.

\subsection{Data} \label{sec:sim-design}

We simulated $G = 3{,}000$ districts with different means and SDs. District means and SDs were drawn independently, with means coming from a uniform distribution ranging from -0.5 to 0.5 and SDs coming from a uniform distribution ranging from 0.5 to 1.5. We regarded each district as having a very large population from which we could draw small, medium, or large samples. For one-third of districts we drew small samples with $n_g=25$ scores per district; for one-third we drew mid-sized samples with $n_g=50$ scores per district; and for one-third we drew large samples with $n_g=100$ scores per district. Within each district, scores were sampled from a normal distribution, then categorized into $K=4$ bins with cuts at $Y=-2/3,0,2/3$, which were the quantiles of the pooled distribution.

\subsection{Maximum likelihood point estimates}
We compared fast HETOP's binned ML estimates to unbinned estimates calculated from the underlying samples of individual test scores. 

\subsubsection{District means} 
For district means, binned ML estimates were excellent. The correlation between binned and unbinned mean estimates was $0.95$ in small samples and $0.99$ in large ones (Table~\ref{tab:sim-accuracy}). This was true whether cuts were fixed or estimated.

\begin{table}[!ht]
	\centering \small
	%% No threeparttable here: it defers \caption so it can size it to the
	%% first tabular, which pushes the main caption below sub-caption 4a.
	%% With sub-tables the Note goes in a plain full-width minipage.
	\caption{Accuracy of fast HETOP estimates from binned counts}
	\label{tab:sim-accuracy}
	
	\begin{subtable}{\textwidth}
		\centering
		\caption{With fixed cuts}
		\label{tab:sim-accuracy-fixed}
		\setlength{\tabcolsep}{2pt}\scriptsize
		\begin{tabular*}{\textwidth}{@{\extracolsep{\fill}}lrrrrrrrrrrrrrr}
			\toprule
			& \multicolumn{7}{c}{Mean} & \multicolumn{7}{c}{SD} \\
			\cmidrule(lr){2-8} \cmidrule(l){9-15}
			& \multicolumn{4}{c}{Maximum likelihood} & \multicolumn{3}{c}{Empirical Bayes} & \multicolumn{4}{c}{Maximum likelihood} & \multicolumn{3}{c}{Empirical Bayes} \\
			\cmidrule(lr){2-5} \cmidrule(lr){6-8} \cmidrule(lr){9-12} \cmidrule(l){13-15}
			Scores per district & Corr. & Bias & RMSE & FMI & Corr. & Bias & RMSE & Corr. & Bias & RMSE & FMI & Corr. & Bias & RMSE \\
			\midrule
			$25$          & $0.95$ & $0.00$ & $0.13$ & $0.19$ & $0.96$ & $0.00$ & $0.12$ & $0.77$ & $0.07$ & $0.32$ & $0.65$ & $0.85$ & $-0.03$ & $0.17$ \\
			$50$          & $0.97$ & $0.00$ & $0.08$ & $0.18$ & $0.98$ & $0.00$ & $0.07$ & $0.88$ & $0.03$ & $0.18$ & $0.65$ & $0.91$ & $-0.02$ & $0.13$ \\
			$100$         & $0.99$ & $0.00$ & $0.05$ & $0.18$ & $0.99$ & $0.00$ & $0.05$ & $0.93$ & $0.01$ & $0.12$ & $0.65$ & $0.94$ & $-0.01$ & $0.10$ \\
			\midrule
			All districts & $0.96$ & $0.00$ & $0.10$ & $0.18$ & $0.97$ & $0.00$ & $0.08$ & $0.83$ & $0.04$ & $0.22$ & $0.65$ & $0.90$ & $-0.02$ & $0.14$ \\
			\bottomrule
		\end{tabular*}
	\end{subtable}
	
	\vspace{1.5em}
	
	\begin{subtable}{\textwidth}
		\centering
		\caption{With estimated cuts}
		\label{tab:sim-accuracy-est}
		\setlength{\tabcolsep}{2pt}\scriptsize
		\begin{tabular*}{\textwidth}{@{\extracolsep{\fill}}lrrrrrrrrrrrrrr}
			\toprule
			& \multicolumn{7}{c}{Mean} & \multicolumn{7}{c}{SD} \\
			\cmidrule(lr){2-8} \cmidrule(l){9-15}
			& \multicolumn{4}{c}{Maximum likelihood} & \multicolumn{3}{c}{Empirical Bayes} & \multicolumn{4}{c}{Maximum likelihood} & \multicolumn{3}{c}{Empirical Bayes} \\
			\cmidrule(lr){2-5} \cmidrule(lr){6-8} \cmidrule(lr){9-12} \cmidrule(l){13-15}
			Scores per district & Corr. & Bias & RMSE & FMI & Corr. & Bias & RMSE & Corr. & Bias & RMSE & FMI & Corr. & Bias & RMSE \\
			\midrule
			$25$          & $0.95$ & $0.00$ & $0.13$ & $0.19$ & $0.96$ & $0.00$ & $0.11$ & $0.77$ & $0.03$ & $0.30$ & $0.65$ & $0.85$ & $-0.01$ & $0.17$ \\
			$50$          & $0.97$ & $0.00$ & $0.08$ & $0.18$ & $0.98$ & $0.00$ & $0.07$ & $0.88$ & $-0.01$ & $0.17$ & $0.65$ & $0.91$ & $0.00$ & $0.13$ \\
			$100$         & $0.99$ & $0.00$ & $0.05$ & $0.18$ & $0.99$ & $0.00$ & $0.05$ & $0.93$ & $-0.03$ & $0.12$ & $0.65$ & $0.94$ & $0.01$ & $0.10$ \\
			\midrule
			All districts & $0.96$ & $0.00$ & $0.09$ & $0.18$ & $0.97$ & $0.00$ & $0.08$ & $0.83$ & $0.00$ & $0.21$ & $0.65$ & $0.90$ & $0.00$ & $0.13$ \\
			\bottomrule
		\end{tabular*}
	\end{subtable}
	
	\vspace{0.75ex}
	\begin{minipage}{\textwidth}\footnotesize
		\textit{Note.} Binned estimates are compared to the means and SDs of
		the unbinned sample scores. Corr.\ is the correlation between
		binned and unbinned estimates. Bias is average error and RMSE is
		root mean square error, both in units of the total SD of the
		pooled score distribution. FMI is the median fraction of missing information
		lost to binning.
	\end{minipage}
\end{table}
\FloatBarrier

\begin{figure}[!ht]
	\centering
	\includegraphics[width=\textwidth]{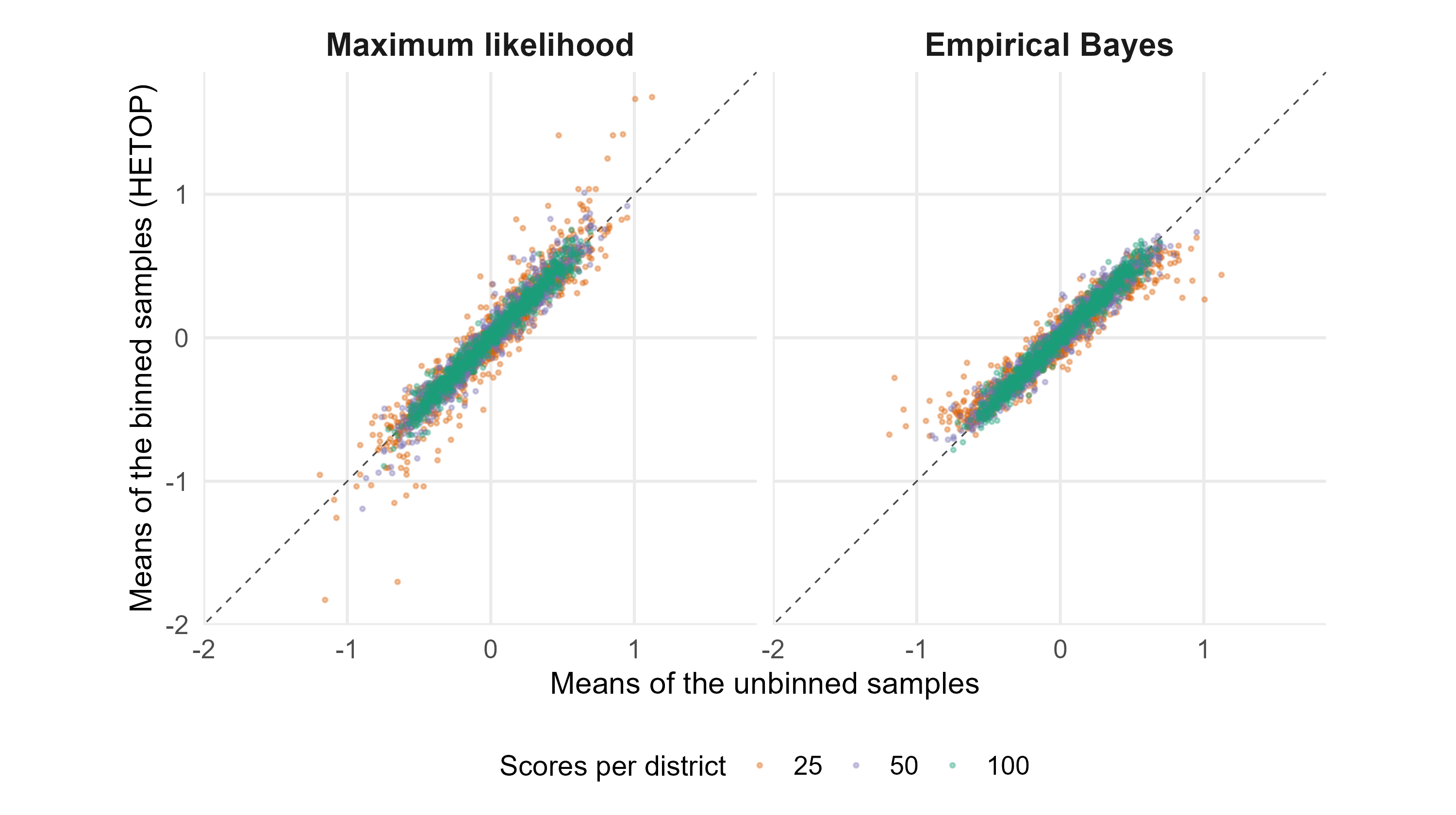}\\[1ex]
	\includegraphics[width=\textwidth]{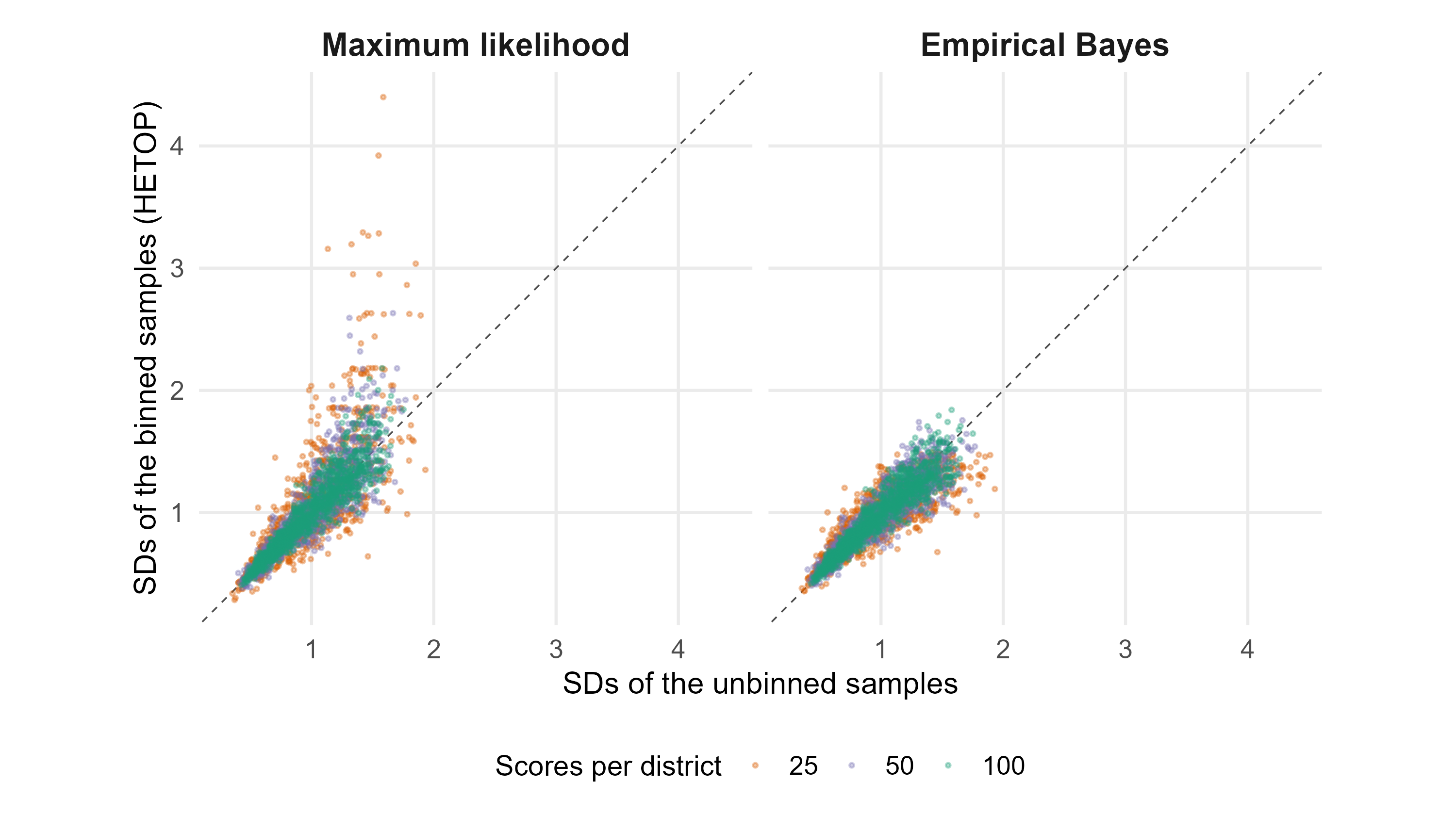}
	\caption{$Y$: District means and SDs, estimated from binned counts by fast HETOP with fixed cuts. $X$: District means and SDs estimated from unbinned sample scores. Top: district means. Bottom: district SDs. Left: Maximum likelihood estimates. Right: empirical Bayes. Perfectly accurate estimates would lie on the diagonal ($Y = X$).}
	\label{fig:sim-scatter}
\end{figure}

\begin{figure}[!ht]
	\centering
	\includegraphics[width=\textwidth]{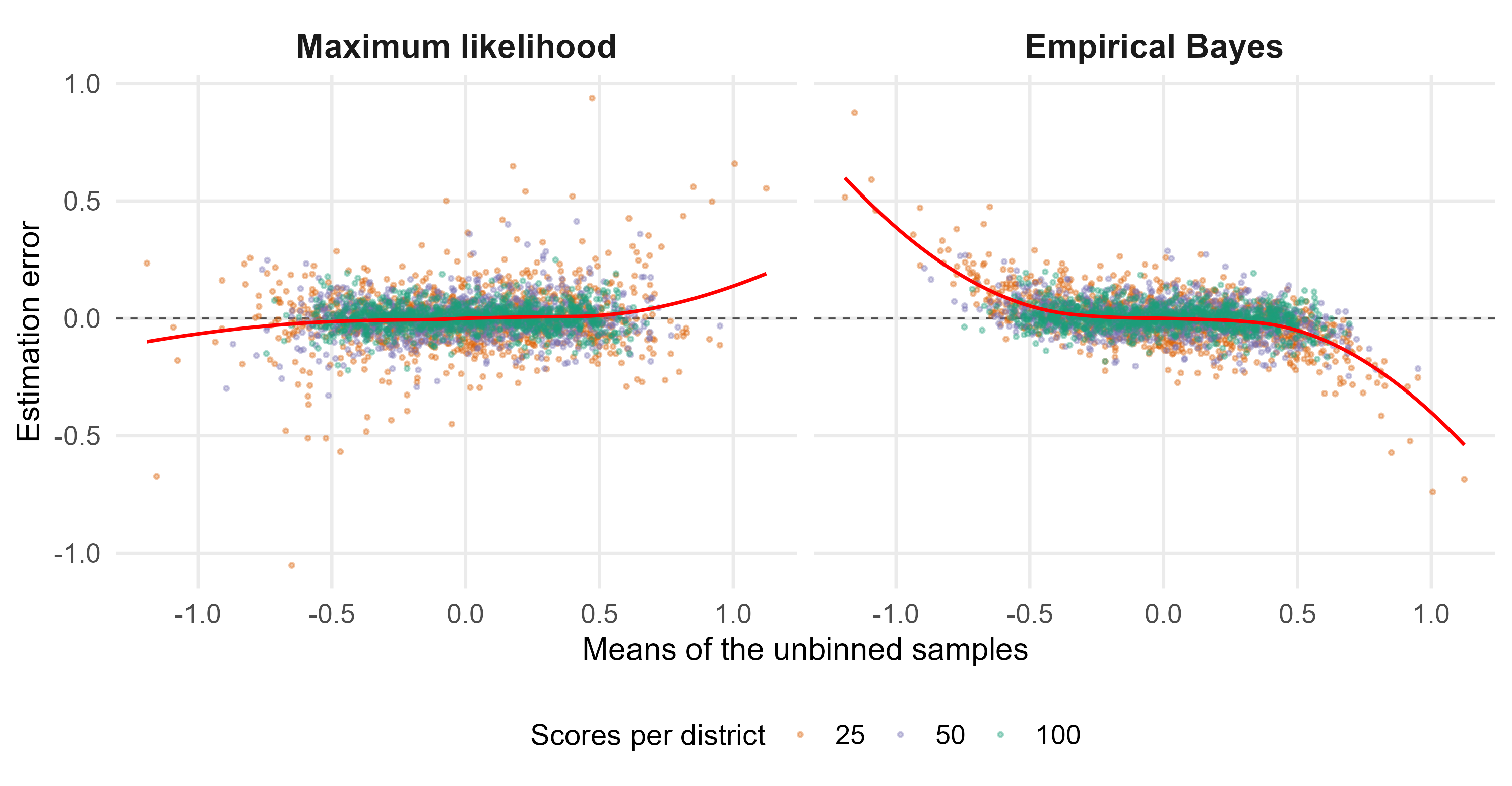}\\[1ex]
	\includegraphics[width=\textwidth]{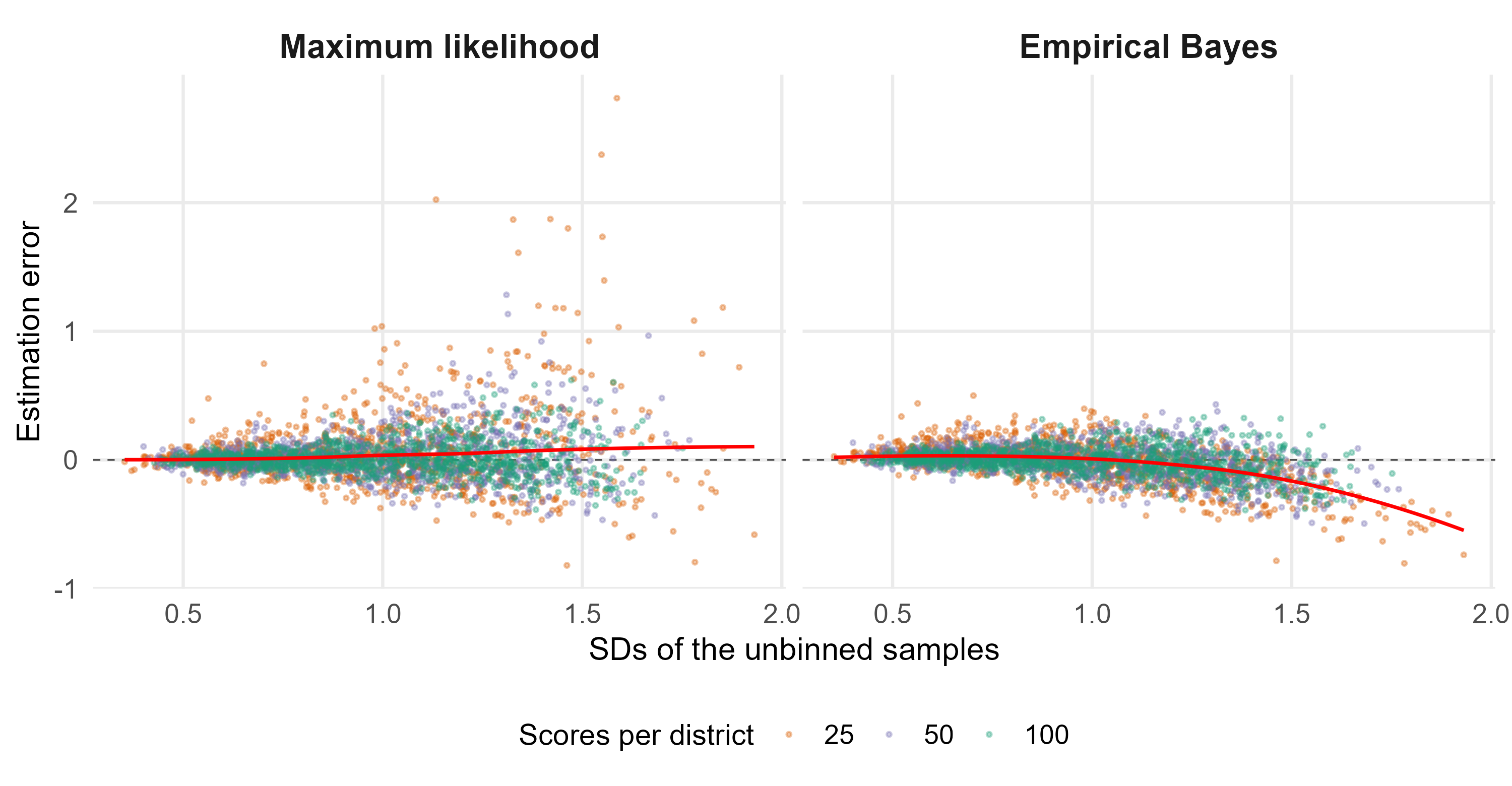}
	\caption{$Y$: Estimation error of district means and SDs, estimated from binned counts by fast HETOP with fixed cuts. $X$: District means and SDs estimated from unbinned sample scores. Top: district means. Bottom: district SDs. Left: Maximum likelihood estimates. Right: empirical Bayes. Unbiased estimates would scatter evenly about zero.}
	\label{fig:sim-error}
\end{figure}

Mean estimates were unbiased in the middle of the distribution, but biased at the extremes. The bias was positive (too high) when the true mean was high, and negative (too low) when the true mean was low. The positive and negative biases canceled each other, so that overall the bias in estimating the mean was zero (Table~\ref{tab:sim-accuracy}). The positive and negative biases are just noticeable when we scatterplot the estimated means (Figure~\ref{fig:sim-scatter}), and the biases become obvious when we scatterplot the estimation error (Figure~\ref{fig:sim-error}).

\FloatBarrier

These biases have an intuitive explanation. As the proportion of students scoring in the top bin increases, ML estimation tries to maximize the likelihood that students will score in the top bin and minimize the likelihood that they will score in lower bins. In the extreme, if all students score in the top bin, which is unbounded above, the ML estimate of the mean becomes unbounded itself, flying off to $+\infty$ (Appendix~\ref{app:unident}). In a less extreme district, where \emph{nearly} all students score in the highest bin, we see a hint of the estimate taking flight---resulting in a positive bias. In a similar way, we can explain the negative bias in the lowest-scoring districts where scores are concentrated in the unbounded bottom bin.

\subsubsection{District SDs} 
For district SDs, estimates were not as good; the correlation between binned and unbinned estimates was $0.93$ in large samples but only $0.77$ in small ones. The relative inefficiency of the SD estimates is visible in Figure~\ref{fig:sim-scatter}, where the cloud of estimated vs true values is more scattered for the SD than for the mean. The SD cloud also shows some outliers---extreme overestimates, concentrated in small districts where the true SD was large. These overestimates contributed to a positive bias, equal to 7\% of the total SD in small districts, and shrinking to 1\% in large districts. Using estimated cuts reduced small-sample bias in the SD estimates, but did not eliminate the outliers.

The lower quality of the binned SD estimates is reflected in the fraction of missing information (FMI), which has a median of 65\% for the SD vs just 18\% for the mean (Table~\ref{tab:sim-accuracy}). The FMI, defined in Appendix~\ref{app:releff}, represents the fraction of information about each parameter lost to binning. The amount by which binning increases the SE---i.e., the ratio of the binned SE to the SE that would have been obtained from unbinned scores---is $1/\sqrt{1-\mathrm{FMI}}$. So in our simulation the median increase in SE due to binning was 69\% for SE estimates but only 11\% for mean estimates.

In Texas, too, binning destroyed more information about the SD than about the mean. The median FMI in Texas was $53\%$ for estimated SDs vs just $16\%$ for estimated means.\footnote{We can calculate the FMI for SD estimates in Texas even though we lack real SD values to compare to the estimates. Calculating FMI involves comparing the SE of estimates from binned and hypothetical unbinned data. It does not require knowing the true values of the parameters being estimated.} It is not always true that the FMI is greater for the SD, but it seems to be true in most realistic data.

\subsection{Empirical Bayes point estimates}
Empirical Bayes shrinks low-information estimates toward the grand mean. Shrinkage generally improved the estimates. It improved estimates of the SD more than estimates of the mean, and it improved estimates more in small samples than in large ones.  

\subsubsection{District means}
For district means, EB reduced excess variability, reduced mean squared error, and increased the correlation between binned and unbinned estimates --- but only slightly. EB underestimated the means of high-scoring districts (positive bias) and overestimated the means of low-scoring districts (negative bias) --- reversing the pattern of bias that we noticed in the ML estimates. The positive and negative biases canceled, so that the overall bias was 0.

\FloatBarrier

\subsubsection{District SDs}
For district SDs, EB pulled in outlying SD estimates (Figure~\ref{fig:sim-scatter}) and increased the correlation between binned and unbinned estimates (Table~\ref{tab:sim-accuracy}). In high-SD districts, EB estimates had a slight negative bias --- reversing the positive bias of the ML estimates.

\subsection{Standard errors and confidence intervals} \label{sec:sim-coverage}

Table~\ref{tab:sim-coverage} shows that confidence intervals have approximately 95\% coverage in the simulated normal data. This confirms that confidence intervals do have nominal coverage when the normal assumption of the HETOP model is satisfied. It was non-normality that caused the low coverage of confidence intervals in Texas.

Half the confidence intervals address only binning error and cover only the sample mean that would have been obtained if scores were not binned. The other intervals, which are longer, account for both binning and sampling, so that they cover the mean of the population from which scores were sampled. Both types of interval have close to 95\% coverage, meaning they cover the quantity they are designed to cover in approximately 95\% of districts.

EB intervals cover just as well as ML intervals, even though EB intervals are shorter due to shrinkage. The intervals cover the SD as well as they cover the mean, and they cover equally well with fixed cuts or estimated cuts.

\begin{table}[!ht]
	\centering \small
	\begin{threeparttable}
	\caption{Coverage and median width of nominal $95\%$ CIs in $3{,}000$ simulated districts}
	\label{tab:sim-coverage}
	\begin{tabular}{lllrrrr}
		\toprule
		& & & \multicolumn{2}{c}{Maximum likelihood} & \multicolumn{2}{c}{Empirical Bayes} \\
		\cmidrule(lr){4-5} \cmidrule(l){6-7}
		Source of error addressed & Quantity covered & Cuts & Width & Coverage & Width & Coverage \\
		\midrule
		Binning & Sample mean    & fixed     & $0.25$ & $96\%$ & $0.25$ & $94\%$ \\
		Binning & Sample mean    & estimated & $0.24$ & $94\%$ & $0.25$ & $95\%$ \\
		Binning+Sampling & Population mean & fixed     & $0.58$ & $95\%$ & $0.52$ & $95\%$ \\
		Binning+Sampling & Population mean & estimated & $0.56$ & $95\%$ & $0.54$ & $95\%$ \\
		\midrule
		Binning   & Sample SD    & fixed     & $0.51$ & $96\%$ & $0.47$ & $95\%$ \\
		Binning   & Sample SD    & estimated & $0.49$ & $93\%$ & $0.48$ & $95\%$ \\
		Binning+Sampling   & Population SD & fixed     & $0.64$ & $95\%$ & $0.57$ & $95\%$ \\
		Binning+Sampling   & Population SD & estimated & $0.62$ & $93\%$ & $0.59$ & $96\%$ \\
		\bottomrule
	\end{tabular}
	\begin{tablenotes}[flushleft]\footnotesize
		\item \textit{Note.}  Widths are in units of the total SD of the pooled score distribution.
	\end{tablenotes}
	\end{threeparttable}
\end{table}
\FloatBarrier

\section{Why is fast HETOP fast?} \label{sec:methods}

Fast HETOP is fast because it breaks the estimation problem down into small pieces, each of which can be solved with just a handful of parameters and little or no iteration.

\subsection{One district, cuts fixed} \label{sec:one-district}

Let's start with the simplest case: a single district with fixed cuts and scores in at least three bins. Then the district mean $\mu_g$ and SD $\sigma_g$ can be estimated using maximum likelihood. The likelihood $L_g(\mu_g, \sigma_g)$ can be written in closed form, and so can its first and second derivatives. Using these formulas, ML point estimates and SEs can be obtained with minimal iteration---in fact, no iteration at all if there are only three bins. Appendix~\ref{app:ml} derives the formulas and solutions, which were first derived in the 1940s and 1950s, when they had to be solved by hand \citep{gjeddebaek1949,stevens1948,gjeddebaek1956,kulldorff1958,kulldorff1958b}.

\subsection{Many districts, cuts fixed} \label{sec:many-districts}

Now suppose there are many districts, all sharing the same fixed cuts. Here the key insight is that districts are independent; that is, each district's mean and SD can be estimated from its own bin counts, ignoring other districts. So the formulas for one district can be applied to each district individually. 

Put more technically, the joint likelihood for all districts together can be factored into the product of each district's individual likelihood:
\begin{equation}
	L(\mu_1, \sigma_1, \dots, \mu_G, \sigma_G) \;=\; \prod_{g=1}^{G} L_g(\mu_g, \sigma_g) .
	\label{eq:lik-factor}
\end{equation}
\noindent So maximizing each district's individual likelihood maximizes the joint likelihood as well. 

\subsection{Many districts, cuts estimated} \label{sec:est-cuts}
Next suppose the shared cuts need to be estimated along with the district means and SDs. Now the district likelihoods are no longer independent, but interconnected through the cuts. That is, moving the cuts affects the likelihood of every district's mean and SD, and moving the district means and SDs affects the likelihood of the cuts.

This makes estimation harder, but we can simplify it by starting with guesses for the cuts. Then we can iterate the following steps until convergence:

\begin{enumerate}
\item Treat the cuts as fixed, then conditionally estimate every district's mean and SD independently by maximum likelihood.
\item Treat the district means and SDs as fixed, then conditionally re-estimate the cuts by maximum likelihood.
\end{enumerate}

This approach --- alternating maximization between two blocks of parameters, with each block treating the other as fixed --- is known as \emph{block coordinate ascent} \citep{wright2015}. It converges to maximum likelihood estimates, and convergence can be rapid if the initial guess for the cuts is close. In our data---both simulated and Texas---the method converged in 2 or 3 iterations. Details of the approach appear in Appendix~\ref{app:cuts}.

\subsection{Empirical Bayes shrinkage} \label{sec:fast-eb}
To get empirical Bayes estimates, we shrink the ML estimates toward their grand mean. We calculate the shrinkage factor using simple, non-iterative formulas (Appendix~\ref{app:eb}), which increase total runtime by only about 10 percent---e.g., from 0.35 to 0.39 seconds for the 1,134 districts in Texas.

\section{Why is joint HETOP slow?} \label{sec:why-slow}

Joint HETOP algorithms are slower. There are three joint HETOP programs and each is slow for a different reason, so we will discuss them separately.

\subsection{Joint HETOP with maximum likelihood}
\paragraph{Large, sparse matrices.} Joint HETOP maximizes the likelihood of all districts at once, rather than one district at a time. If there are $G$ districts, joint HETOP must estimate a vector of $2G$ district parameters --- $G$ means and $G$ standard deviations --- along with its $2G \times 2G$ information matrix, which it must invert to make the variance-covariance matrix of the estimates. For example, if there are $G=1000$ districts, then the variance-covariance matrix has 4 million cells, which can cause memory and computational problems in some environments. Stata/BE, for example, will not accept a matrix larger than $800 \times 800$; the less restricted SE and MP flavors allow $11{,}000 \times 11{,}000$ and $65{,}534 \times 65{,}534$ \citep{statacorp2026limits}. Even where the matrix fits, inverting a large information matrix is computationally expensive: the number of operations required is $O(G^3)$ \citep{golub2013}.

If cuts are fixed, though, the vast majority of cells in the variance-covariance matrix are zero, because they represent covariances between estimates from different districts---which are independent, as we've pointed out (Section~\ref{sec:many-districts}). Again, if there are $G=1000$ districts, then the variance-covariance matrix has 4 million cells --- \emph{but 99.9 percent of them are zeroes}. Only 4,000 cells (0.1 percent) represent within-district variances and covariances that actually need to be estimated. Existing joint HETOP software does not exploit this and estimates the value of every cell---estimating 1,000 times more matrix cells than it needs to. Fast HETOP, by contrast, estimates one 2x2 information matrix for each district -- just 4,000 matrix cells in all.

If cuts are estimated, then estimates from different districts are correlated, but they are still independent when conditioned on the cuts. Fast HETOP exploits this conditional independence (Section~\ref{sec:est-cuts}). Joint HETOP does not.

This is the main reason that joint HETOP software is slow when the number of districts is large.

\paragraph{Numerical derivatives.} Another reason joint HETOP is slow is that it relies on numerical methods. Finding the maximum of the likelihood requires its derivatives. Fast HETOP uses specific, explicit formulas for the first and second derivatives (Appendix~\ref{app:ml}), while joint HETOP approximates the derivatives with general-purpose numeric methods. Both approaches can maximize the likelihood, but explicit formulas get there faster. Using formulas is like having a contour map, while even the best numeric methods are a bit like groping the ground.

\subsection{Bayesian joint HETOP}
\paragraph{Markov chain Monte Carlo.} Bayesian HETOP is slow for a different reason. Instead of maximizing the likelihood, it uses Markov Chain Monte Carlo (MCMC) to sample from the posterior distribution of the district means and SDs \citep{lockwood2018}. MCMC is slow because it samples hundreds or thousands of values for each district, and ``burns'' hundreds or thousands more---discarding early draws until the chain converges on the posterior, and intermediate draws so that the values it keeps are nearly independent.

MCMC can be the best option for difficult estimation problems that cannot be solved in other ways. For the binned data problem, though, fast HETOP can produce very similar estimates much more quickly using simple formulas that require no sampling (Appendix~\ref{app:eb}).

\section{Conclusion} \label{sec:discussion}

The fast HETOP algorithm is a much faster way to get ML estimates of district means and SDs from binned data under a normal model. Mean estimates are strongly correlated with the means we would get if the data were not binned, but SD estimates are not as good. Maximum likelihood estimates from binned data can be biased or unreliable, with poor coverage, if scores are non-normal or concentrated in the lowest or highest bins. Empirical Bayes methods can substantially improve unreliable estimates and should be used routinely.

While the estimates that can be extracted from binned data are surprisingly good, they are not perfect, and their limitations are fundamental. Binning test scores destroys information, especially about the SD, and there is no statistical method that can restore all the information that has been lost. It would be much better if education agencies reported means and SDs for each district and school directly, and there is no plausible reason not to. Until all agencies report means and SDs directly, though, fast HETOP offers researchers a convenient way to get the data they need to estimate trends and effects --- only approximately, but well enough for many purposes.

%% References------------------------------------------------------
\label{page:refstart}
\bibliography{references}

%% Appendix--------------------------------------------------------
%% JREE wants references before the appendices.
\appendix
\label{page:appendixstart}
\numberwithin{equation}{section}   %% appendix equations become A.1, A.2, ..., B.1, ...

%% One "Appendices" heading covers them all, so the individual titles
%% carry only their letter.  Prefixing each title with "Appendix" made
%% the longer ones wrap and hyphenate across two lines.
\section*{Appendices}

\section{Maximum likelihood estimation of district means and SDs} \label{app:ml}

This appendix describes how we maximized the likelihood of district means and SDs.

Within district $g$, $n_g$ students take a test and receive scores $Y$, which we assume to be normally distributed with mean $\mu_g$ and SD $\sigma_g$. We cannot observe the individual scores, but only observe how many students $n_{gk}$ scored in bins $k=1,..,K$, where the $K$ bins are delimited by $K-1$ cuts $c_1 < c_2 < \dots < c_{K-1}$. 

We assume throughout this Appendix that the cuts are known or \emph{fixed}. Appendix~\ref{app:cuts} generalizes to the case where cuts must be estimated.

\subsection{Key quantities} \label{app:binned-data}

Our goal is to get point estimates of $\mu_g$ and SD $\sigma_g$. 

\paragraph{Observed proportions}Our estimates will reconcile, as closely as possible, the proportion of scores $p$ observed in each bin with the bin probabilities $P$ expected according to a fitted normal distribution. The observed proportion of scores in each bin $k$ is 
\begin{equation}
p_{gk} = \frac{n_{gk}}{n_g}
	\label{eq:p}
\end{equation}

\noindent and the cumulative proportion of scores in the bottom $k$ bins is
\begin{equation}
	\hat F_{gk} = \sum_{j \le k} p_{gj},
	\label{eq:Fhat}
\end{equation}

\paragraph{Modeled probabilities}If scores are normally distributed, the probability of scoring in the bottom $k$ bins is the cumulative probability of scoring below cut $c_k$:
\begin{equation}
	F_{gk}(\mu_g, \sigma_g) = \Phi\!\left(\frac{c_{k}-\mu_{g}}{\sigma_{g}}\right) = \Phi(z_{gk}),
	\label{eq:F}
\end{equation}

\noindent where $\Phi$ is the standard-normal cumulative distribution function (CDF) and $z_{gk}=(c_{k}-\mu_{g})/\sigma_{g}$ is the cut score standardized by the district mean and SD. Then the probability of scoring in bin $k$ is
\begin{equation}
	P_{gk}(\mu_g, \sigma_g) = F_{gk} - F_{g(k-1)} = \Phi(z_{gk}) - \Phi(z_{g(k-1)}), k=1,...,K
	\label{eq:P}
\end{equation}

\noindent Since the bottom and top bins are unbounded, the bottom cut is $z_{g0}=-\infty$ with CDF $F_{g0}=0$, and the top cut is $z_{gK}=+\infty$ with CDF $F_{gK}=1$.

\subsection{The likelihood} \label{app:likelihood}

We estimate the district means and SDs by maximum likelihood. Under a normal model with district mean $\mu_g$ and district SD $\sigma_g$, the likelihood of observing $n_{gk}$ scores in bin $k=1,...K$ is 
\begin{equation}
	L_g(\mu_g, \sigma_g) \;\propto\; \prod_{k=1}^K \bigl(P_{gk}(\mu_g, \sigma_g)\bigr)^{\,n_{gk}},
	\label{eq:lik}
\end{equation}

\noindent and the log likelihood is
\begin{equation}
	\ell_g(\mu_g, \sigma_g) \;=\; \log L_g(\mu_g, \sigma_g) \;=\; n_g \sum_{k=1}^K p_{gk} \log P_{gk}(\mu_g, \sigma_g),
	\label{eq:loglik}
\end{equation}

\noindent The ML estimates are the values $(\hat\mu_g, \hat\sigma_g)$ that maximize the log-likelihood. 

\subsection{Maximizing the likelihood with 3 bins requires no iteration} \label{app:three-bins}

The 2 parameters $(\mu_g, \sigma_g)$ are \emph{identified} if there are 3 or more bins populated with counts greater than zero. With exactly $K=3$ bins, there are 2 cuts, and we can maximize the likelihood just by setting the modeled cumulative probability equal to the observed cumulative proportion at each cut:
\begin{align}
	F_{g1} &\;=\; \hat F_{g1},
	\label{eq:Fhat-F-1} \\
	F_{g2} &\;=\; \hat F_{g2},
	\label{eq:Fhat-F-2}
\end{align}

\noindent Since $F_{gk} = \Phi(z_{gk})$, setting $F_{gk}=\hat F_{gk}$ implies $z_{gk} = \Phi^{-1}(\hat F_{gk})$, and since $z_{gk}=(c_k-\mu_g)/\sigma_g$, our two equations can be re-expressed as
\begin{align}
	c_1 &\;=\; \hat\mu_g + \hat\sigma_g\,\Phi^{-1}(\hat F_{g1}),
	\label{eq:cut-quantile-1} \\
	c_2 &\;=\; \hat\mu_g + \hat\sigma_g\,\Phi^{-1}(\hat F_{g2}).
	\label{eq:cut-quantile-2}
\end{align}
\noindent Subtracting Equation~\eqref{eq:cut-quantile-1} from Equation~\eqref{eq:cut-quantile-2} and rearranging, we get
\begin{equation}
	\hat\sigma_g = \frac{c_2 - c_1}{\Phi^{-1}(\hat F_{g2}) - \Phi^{-1}(\hat F_{g1})},
	\label{eq:sigma_g_k3}
\end{equation}
\noindent and substituting $\hat\sigma_g$ back into Equation~\eqref{eq:cut-quantile-1} gives
\begin{equation}
	\hat\mu_g = c_1 - \hat\sigma_g \Phi^{-1}(\hat F_{g1}).
	\label{eq:mu_g_k3}
\end{equation}
$\hat\mu_g$ and $\hat\sigma_g$ are ML estimates %% The self-citation is withheld from the anonymous version; the
%% remaining three sources establish the result on their own.
\ifanon\citep{gjeddebaek1949,stevens1948,gjeddebaek1959b}\else\citep{gjeddebaek1949,stevens1948,gjeddebaek1959b,vonhippel2019}\fi and can be calculated without iteration.

With 3 bins and 2 cuts, the district normal model will fit the observed proportions exactly. This does not mean that the normal model is correct. It just means that the number of cuts is equal to the number of parameters. Whatever the underlying test score distribution looks like, a normal model with 2 parameters can always reproduce the observed cumulative proportions at 2 cuts.

\subsection{Maximizing the likelihood with 4 bins or more} \label{app:kge4}

If there are more than $K=3$ bins, the normal model cannot, in general, fit the bin counts perfectly, so we cannot maximize the likelihood simply by setting the model-based bin proportions $P_{gk}$ equal to the observed proportions $p_{gk}$. Instead, maximization requires differentiating the log likelihood.

%% Condensed for the page-limited (anonymous) journal version; the
%% full derivation is kept for the working paper.
\ifanon
\subsubsection{Differentiating the log likelihood}

Differentiating the log likelihood with respect to $\mu_g$ and $\sigma_g$ requires the chain rule, applied first to the logarithm and then to $P_{gk}$, which is a difference of two normal CDFs. The result is \citep{stevens1948,gjeddebaek1956}
\begin{align}
	\frac{\partial \ell_g}{\partial \mu_g}
	&= \frac{n_g}{\sigma_g} \sum_{k=1}^K p_{gk}\,
	  \frac{\varphi(z_{g(k-1)}) - \varphi(z_{gk})}{P_{gk}},
	\label{eq:app-raw-mu} \\[4pt]
	\frac{\partial \ell_g}{\partial \sigma_g}
	&= \frac{n_g}{\sigma_g} \sum_{k=1}^K p_{gk}\,
	  \frac{z_{g(k-1)}\varphi(z_{g(k-1)}) - z_{gk}\varphi(z_{gk})}{P_{gk}}.
	\label{eq:app-raw-sigma}
\end{align}

\else
\subsubsection{Differentiating the log likelihood} \label{app:score-eqs}

 Differentiating the log likelihood requires applying the chain rule twice. The first application turns the logarithm into a reciprocal:
\begin{align}
	\frac{\partial \ell_g}{\partial \mu_g}	
= n_g \sum_{k=1}^K \frac{p_{gk}}{P_{gk}}\,\frac{\partial P_{gk}}{\partial \mu_g},
	\label{eq:app-chain1} \\
	\frac{\partial \ell_g}{\partial \sigma_g}	
= n_g \sum_{k=1}^K \frac{p_{gk}}{P_{gk}}\,\frac{\partial P_{gk}}{\partial \sigma_g},
\label{eq:app-chain1b}	
\end{align}

\noindent The second application of the chain rule involves differentiating $P_{gk}$. We expand $P_{gk}$ into a difference of two CDFs, then differentiate each CDF, using the fact that the derivative of the normal CDF $\Phi$ is the normal PDF $\varphi$:
\begin{align}
	\frac{\partial P_{gk}}{\partial \mu_g}
	&= \frac{\partial}{\partial \mu_g}\bigl[\Phi(z_{gk}) - \Phi(z_{g(k-1)})\bigr]
	\nonumber \\[4pt]
	&= \varphi(z_{gk})\,\frac{\partial z_{gk}}{\partial \mu_g}
	 - \varphi(z_{g(k-1)})\,\frac{\partial z_{g(k-1)}}{\partial \mu_g},
	\label{eq:app-dPmu-chain} \\[10pt]
	\frac{\partial P_{gk}}{\partial \sigma_g}
	&= \frac{\partial}{\partial \sigma_g}\bigl[\Phi(z_{gk}) - \Phi(z_{g(k-1)})\bigr]
	\nonumber \\[4pt]
	&= \varphi(z_{gk})\,\frac{\partial z_{gk}}{\partial \sigma_g}
	 - \varphi(z_{g(k-1)})\,\frac{\partial z_{g(k-1)}}{\partial \sigma_g}.
	\label{eq:app-dPsigma-chain}
\end{align}

Since $z_{gk} = (c_k - \mu_g)/\sigma_g$, the $z_{gk}$'s have derivatives
\begin{equation}
	\frac{\partial z_{gk}}{\partial \mu_g} = -\frac{1}{\sigma_g},
	\qquad
	\frac{\partial z_{gk}}{\partial \sigma_g} = -\frac{z_{gk}}{\sigma_g}.
	\label{eq:app-dz}
\end{equation}

Substituting these into \eqref{eq:app-dPmu-chain} and \eqref{eq:app-dPsigma-chain} gives
\begin{align}
	\frac{\partial P_{gk}}{\partial \mu_g}
	&= \varphi(z_{gk})\left(-\frac{1}{\sigma_g}\right)
	 - \varphi(z_{g(k-1)})\left(-\frac{1}{\sigma_g}\right)
	\nonumber \\[4pt]
	&= \frac{\varphi(z_{g(k-1)}) - \varphi(z_{gk})}{\sigma_g},
	\label{eq:app-dPmu} \\[10pt]
	\frac{\partial P_{gk}}{\partial \sigma_g}
	&= \varphi(z_{gk})\left(-\frac{z_{gk}}{\sigma_g}\right)
	 - \varphi(z_{g(k-1)})\left(-\frac{z_{g(k-1)}}{\sigma_g}\right)
	\nonumber \\[4pt]
	&= \frac{z_{g(k-1)}\varphi(z_{g(k-1)}) - z_{gk}\varphi(z_{gk})}{\sigma_g}.
	\label{eq:app-dPsigma}
\end{align}

\noindent And substituting
Equations~\eqref{eq:app-dPmu} and \eqref{eq:app-dPsigma} into 
Equations~\eqref{eq:app-chain1} and \eqref{eq:app-chain1b} gives the derivatives of the log likelihood:
\begin{align}
	\frac{\partial \ell_g}{\partial \mu_g}
	&= \frac{n_g}{\sigma_g} \sum_{k=1}^K p_{gk}\,
	  \frac{\varphi(z_{g(k-1)}) - \varphi(z_{gk})}{P_{gk}},
	\label{eq:app-raw-mu} \\[4pt]
	\frac{\partial \ell_g}{\partial \sigma_g}
	&= \frac{n_g}{\sigma_g} \sum_{k=1}^K p_{gk}\,
	  \frac{z_{g(k-1)}\varphi(z_{g(k-1)}) - z_{gk}\varphi(z_{gk})}{P_{gk}}.
	\label{eq:app-raw-sigma}
\end{align}
\fi

\paragraph{Re-expressing derivatives in terms of bin moments.} The derivatives can be re-expressed more clearly as
\begin{align}
	\frac{\partial \ell_g}{\partial \mu_g}
	&= \frac{n_g}{\sigma_g^{\,2}} \sum_{k=1}^K p_{gk}\,(\mu_{gk} - \mu_g),
	\label{eq:deriv-mu} \\[6pt]
	\frac{\partial \ell_g}{\partial \sigma_g}
	&= \frac{n_g}{\sigma_g^{\,3}} \sum_{k=1}^K p_{gk}\,
	   \bigl[\sigma^2_{gk} - \sigma^2_g + (\mu_{gk} - \mu_g)^2\bigr],
	\label{eq:deriv-sigma}
\end{align}

where
\begin{align}
	\mu_{gk}    &= \mu_g + \sigma_g \,\frac{\varphi(z_{g(k-1)}) - \varphi(z_{gk})}{P_{gk}},
	\label{eq:mugk} \\[6pt]
	\sigma_{gk}^2 &= \sigma_g^2 \left(1 + \frac{z_{g(k-1)}\,\varphi(z_{g(k-1)}) - z_{gk}\,\varphi(z_{gk})}{P_{gk}} - \left(\frac{\varphi(z_{g(k-1)}) - \varphi(z_{gk})}{P_{gk}}\right)^2\right),
	\label{eq:sigmagk}
\end{align}
are the mean and variance of scores within bin $k$ of district $g$. We know these expressions because each bin fits a truncated normal distribution \citep[pp.~156--158]{johnson1994}. 

\subsubsection{Maximum likelihood point estimates} \label{app:fixed-cuts}
%% Condensed for the page-limited (anonymous) journal version; the
%% full derivation is kept for the working paper.
\ifanon
The ML estimates $\hat\mu_g$ and $\hat\sigma_g^2$ are the values of $\mu_g$ and $\sigma_g^2$ that set both derivatives to zero. Setting Equations~\eqref{eq:deriv-mu} and~\eqref{eq:deriv-sigma} to zero and using $\sum_{k=1}^K p_{gk} = 1$ gives
\else
The ML estimates $\hat\mu_g$ and  $\hat\sigma_g^2$ are the values of $\mu$ and $\sigma^2$ that set both derivatives to zero. Setting the first derivative to zero gives
\begin{align}
	\left.\frac{\partial \ell_g}{\partial \mu_g}\right|_{\mu_g = \hat\mu_g,\; \sigma_g = \hat\sigma_g} &= 0 \notag \\
	\sum_{k=1}^K p_{gk}\,(\hat\mu_{gk} - \hat\mu_g) &= 0 \notag \\
	\sum_{k=1}^K p_{gk}\,\hat\mu_{gk} \;-\; \hat\mu_g \sum_{k=1}^K p_{gk} &= 0 \notag \\
	\sum_{k=1}^K p_{gk}\,\hat\mu_{gk} \;-\; \hat\mu_g &= 0
	   \qquad \text{since } \sum_{k=1}^K p_{gk} = 1 \notag \\
	\hat\mu_g &= \sum_{k=1}^K p_{gk}\, \hat\mu_{gk}
	\label{eq:app-solve-mu}
\end{align}
And setting the second derivative to zero gives
\begin{align}
	\left.\frac{\partial \ell_g}{\partial \sigma_g}\right|_{\mu_g = \hat\mu_g,\; \sigma_g = \hat\sigma_g} &= 0 \notag \\
	\sum_{k=1}^K p_{gk}\,\bigl[\hat\sigma^2_{gk} - \hat\sigma^2_g
	    + (\hat\mu_{gk} - \hat\mu_g)^2\bigr] &= 0 \notag \\
	\sum_{k=1}^K p_{gk}\, \hat\sigma^2_{gk}
	    \;-\; \hat\sigma^2_g \sum_{k=1}^K p_{gk}
	    \;+\; \sum_{k=1}^K p_{gk}\,(\hat\mu_{gk} - \hat\mu_g)^2 &= 0 \notag \\
	\sum_{k=1}^K p_{gk}\, \hat\sigma^2_{gk}
	    \;-\; \hat\sigma^2_g
	    \;+\; \sum_{k=1}^K p_{gk}\,(\hat\mu_{gk} - \hat\mu_g)^2 &= 0
	   \qquad \text{since } \sum_{k=1}^K p_{gk} = 1 \notag \\
	\hat\sigma_g^2 &= \sum_{k=1}^K p_{gk}\, \hat\sigma_{gk}^2
	              \;+\; \sum_{k=1}^K p_{gk}\, (\hat\mu_{gk} - \hat\mu_g)^2
	\label{eq:app-solve-sigma}
\end{align}
\fi

\noindent That is, the ML estimates are
\begin{align}
	\hat\mu_g &= \sum_{k=1}^K p_{gk}\, \hat\mu_{gk},
	\label{eq:app-est-mu} \\[8pt]
	\hat\sigma_g^2 &= \underbrace{\sum_{k=1}^K p_{gk}\, \hat\sigma_{gk}^2}_{\text{variance within bins}}
	\;+\; \underbrace{\sum_{k=1}^K p_{gk}\, (\hat\mu_{gk} - \hat\mu_g)^2}_{\text{variance between bins}}.
	\label{eq:app-est-sigma}
\end{align}

\noindent The district mean is the weighted average of the bin means, and the district variance is the variance within bins plus the variance between them. Since the bin means and variances are functions of the district means and variances, the equations implicitly have functions of $\hat\mu_g$ and $\hat\sigma_g$ on both left and right. We have to search for the values of $\hat\mu_g$ and $\hat\sigma_g$ that make the left and right side equal, and if there are more than 3 bins, the search requires iteration.

To get initial estimates, we collapse to three bins, get the 3-bin estimates from Equations~\eqref{eq:sigma_g_k3} and~\eqref{eq:mu_g_k3}, and call them $(\hat\mu_g^{(0)}, \hat\sigma_g^{(0)})$. We then un-collapse the data to $K \geq 4$ bins and update the estimates iteratively:
\begin{align}
\hat\mu_g^{(t)} &= \sum_{k=1}^K p_{gk}\, \hat\mu_{gk}^{(t-1)}
\label{eq:iter-mu} \\
\hat\sigma_g^{2\,(t)} &= \sum_{k=1}^K p_{gk}\, \hat\sigma_{gk}^{2\,(t-1)} \;+\; \sum_{k=1}^K p_{gk}\, (\hat\mu_{gk}^{(t-1)} - \hat\mu_g^{(t)})^2,
\label{eq:iter-sigma}
\end{align}
\noindent \citep{schader1984}. Iteration starts at $t=1$ and continues until the estimates stop moving---that is, until both $|\hat\mu_g^{(t)} - \hat\mu_g^{(t-1)}| / \hat\sigma_g$ and $|\log \hat\sigma_g^{(t)} - \log \hat\sigma_g^{(t-1)}|$ fall below a small tolerance, set by default to $10^{-4}$. Estimates typically converge in a handful of steps for well-populated districts; sparsely-populated districts take longer. 

This algorithm is known as fixed-point iteration. A slightly more complicated algorithm known as Fisher scoring would converge faster \citep{schader1984}, but fixed-point iteration converges in less than a millisecond per district on average, so we judged it good enough.

\subsection{Standard errors and confidence intervals} \label{app:ml-ci}

In large samples, the standard errors of maximum likelihood estimates can be calculated from the expected information matrix, which is a cross-product of the first derivatives of the log likelihood. For binned normal data, this matrix is \citep{stevens1948,gjeddebaek1956}
\begin{align}
	\mathcal{I}_g &= \operatorname{E}\left[
	\begin{pmatrix} \dfrac{\partial \ell_g}{\partial \mu_g} \\[12pt]
	                \dfrac{\partial \ell_g}{\partial \sigma_g} \end{pmatrix}
	\begin{pmatrix} \dfrac{\partial \ell_g}{\partial \mu_g} &
	                \dfrac{\partial \ell_g}{\partial \sigma_g} \end{pmatrix}
	\right]
	\notag \\[8pt]
	&= \operatorname{E}
	\begin{pmatrix}
	  \left(\dfrac{\partial \ell_g}{\partial \mu_g}\right)^{\!2} &
	  \dfrac{\partial \ell_g}{\partial \mu_g}\,\dfrac{\partial \ell_g}{\partial \sigma_g} \\[14pt]
	  \dfrac{\partial \ell_g}{\partial \mu_g}\,\dfrac{\partial \ell_g}{\partial \sigma_g} &
	  \left(\dfrac{\partial \ell_g}{\partial \sigma_g}\right)^{\!2}
	\end{pmatrix}
	&&
	\notag \\[8pt]
	&= \frac{n_g}{\sigma_g^{2}}
	\mathbf{S}, \qquad \mathbf{S} =
	\begin{pmatrix} S_{\mu\mu} & S_{\mu\sigma} \\[6pt] S_{\mu\sigma} & S_{\sigma\sigma} \end{pmatrix}.
	\label{eq:app-info-S}
\end{align}

\noindent The elements of $\mathbf{S}$, which follow from Equations~\eqref{eq:deriv-mu} and~\eqref{eq:deriv-sigma}, are most compactly written using
\begin{equation}
	\Delta\mu_{gk} = \mu_{gk} - \mu_g,
	\qquad
	\Delta\sigma^2_{gk} = \sigma^2_{gk} - \sigma^2_g
	\label{eq:app-deltas}
\end{equation}
\noindent for the differences, within district $g$, between bin $k$'s moments and the whole district's moments. Then
\begin{align}
	S_{\mu\mu} &= \frac{1}{\sigma_g^{\,2}} \sum_{k=1}^K P_{gk}\,(\Delta\mu_{gk})^2,
	\label{eq:app-Smm} \\[6pt]
	S_{\mu\sigma} &= \frac{1}{\sigma_g^{\,3}} \sum_{k=1}^K P_{gk}\,
	  \Delta\mu_{gk}\bigl[\Delta\sigma^2_{gk} + (\Delta\mu_{gk})^2\bigr],
	\label{eq:app-Sms} \\[6pt]
	S_{\sigma\sigma} &= \frac{1}{\sigma_g^{\,4}} \sum_{k=1}^K P_{gk}\,
	  \bigl[\Delta\sigma^2_{gk} + (\Delta\mu_{gk})^2\bigr]^2.
	\label{eq:app-Sss}
\end{align}

Writing $D = \det\mathbf{S} = S_{\mu\mu}S_{\sigma\sigma}-S_{\mu\sigma}^{2}$, the sampling variances are the diagonal elements of the inverse
\begin{equation}
	\mathcal{I}_g^{-1} = \frac{\sigma_g^{2}}{n_g}\,\mathbf{S}^{-1}
	= \frac{\sigma_g^{2}}{n_g D}
	  \begin{pmatrix} S_{\sigma\sigma} & -S_{\mu\sigma} \\[6pt]
	                  -S_{\mu\sigma} & S_{\mu\mu} \end{pmatrix},
	\label{eq:app-info-inv}
\end{equation}
and the standard errors are their square roots. Substituting $(\hat\mu_g, \hat\sigma_g)$ for the unknown $(\mu_g, \sigma_g)$, we estimate the standard errors as
\begin{align}
	\mathrm{SE}(\hat\mu_g)    &= \frac{\hat\sigma_g}{\sqrt{n_g}}\sqrt{\frac{S_{\sigma\sigma}}{D}},
	\label{eq:app-se-mu} \\[4pt]
	\mathrm{SE}(\hat\sigma_g) &= \frac{\hat\sigma_g}{\sqrt{n_g}}\sqrt{\frac{S_{\mu\mu}}{D}}.
	\label{eq:app-se-sigma-raw}
\end{align}

\noindent A $100(1-\alpha)\%$ confidence interval for $\mu_g$ is $\hat\mu_g \pm z_{1-\alpha/2}\,\mathrm{SE}(\hat\mu_g)$, where $z_{1-\alpha/2}$ is the corresponding standard normal quantile.

A confidence interval for $\sigma_g$ is better built on the log scale, since the sampling distribution of $\log\hat\sigma_g$ is more nearly symmetric. By the delta method, the standard error of $\log\hat\sigma_g$ is $1/\sigma_g$ times the standard error of $\hat\sigma_g$, because $1/\sigma_g$ is the derivative of $\log\sigma_g$:
\begin{equation}
	\mathrm{SE}(\log\hat\sigma_g) = \frac{1}{\sqrt{n_g}}\sqrt{\frac{S_{\mu\mu}}{D}}.
	\label{eq:app-se-sigma}
\end{equation}
\noindent A confidence interval for $\log\sigma_g$ is $\log\hat\sigma_g \pm z_{1-\alpha/2}\,\mathrm{SE}(\log\hat\sigma_g)$, and exponentiating the endpoints yields an interval for $\sigma_g$.

\subsection{Districts whose mean and SD are not identified} \label{app:unident}

\subsubsection{Districts with 2 populated bins}
If only 2 of the $K$ bins are populated in a given district, it is not possible to identify that district's mean and SD from the bin counts alone. Yet the bin counts may still be suggestive; for example, the mean score is likely high if counts are concentrated in the upper bins.

A variety of approaches have been tried for unidentified districts \citep{reardon2017,shear2019,shear2021}. We follow Lockwood et al. (\citeyear{lockwood2018}, \citeyear{lockwood2019}), setting the SD of each unidentified district to $\bar\sigma$---the geometric mean of the SDs of the identified districts---and then estimating the district's mean from its bin counts with the SD held at that value. Specifically, we solve for $\hat\mu_g$ by plugging $\bar\sigma$ into equation~\eqref{eq:iter-mu} and then iterating until convergence.

\subsubsection{Districts with 1 populated bin}
If only 1 bin is populated, estimation of the district mean depends on whether the populated bin is an interior bin or an exterior bin.

If the populated bin is interior, with an upper and lower cut, then the likelihood is maximized by putting the mean in the center of that bin:
\begin{equation}
	\hat\mu_g = (c_{j-1} + c_j)/2.
\end{equation}
\noindent which minimizes the probability spilling into the empty neighboring bins.

If the only populated bin is the bottom or top bin, the likelihood cannot be maximized \citep{kulldorff1958,kulldorff1958b}. Again following Lockwood et al. (\citeyear{lockwood2018}, \citeyear{lockwood2019}), we assign such districts the smallest or largest mean estimated for any other district.

\section{Maximum likelihood estimation of cut scores} \label{app:cuts}

If the cuts are unknown (or they are known but we prefer not to use them), cuts can be identified by fixing the mean $\mu$ and SD $\sigma$ for the score distribution of all districts pooled together. These can be the statewide mean and SD, which are reported by some state education agencies (e.g., Texas), or they can be fixed to a mean of 0 and an SD of 1, so that estimates are returned on a standardized scale.

Now we initialize the cuts to plausible starting values, then alternate between the two conditional problems:
\begin{enumerate}
	\item Holding the cuts fixed, maximize the conditional likelihood of each district's mean and SD, using the fixed-cut likelihood of Appendix~\ref{app:ml}.
	\item Holding the district means and SDs fixed, maximize the conditional likelihood of the cuts.
\end{enumerate}
\noindent and repeat until no cut moves by more than a small tolerance, set by default to $10^{-3}\sigma$. Alternating maximization in this way is known as \emph{block coordinate ascent} \citep{wright2015}.

Step 2 in our alternating algorithm requires maximizing the likelihood of the cuts conditionally on the district means and SDs. The conditional likelihood and log likelihood are
\begin{equation}
\begin{aligned}
	L(c_1,\dots,c_{K-1}) &\;\propto\; \prod_{g=1}^G \prod_{k=1}^K \bigl(P_{gk}\bigr)^{\,n_{gk}}, \\[4pt]
	\ell(c_1,\dots,c_{K-1}) &\;=\; \log L(c_1,\dots,c_{K-1}) \;=\; \sum_{g=1}^G \sum_{k=1}^K n_{gk} \log P_{gk}.
\end{aligned}
	\label{eq:cut-loglik}
\end{equation}
\noindent Although the log likelihood uses the same expression as the log likelihood for district means and SDs (Equations~\eqref{eq:lik} and~\eqref{eq:loglik}), considering it as a function of the cuts changes the way it must be maximized. The cuts are shared across districts and their log likelihood cannot be maximized one district at a time. Instead, we iteratively maximize across all districts at once using Newton--Raphson iteration:
\begin{equation}
	\mathbf{c}^{(t+1)} = \mathbf{c}^{(t)} - \left[ \nabla^2 \ell\bigl(\mathbf{c}^{(t)}\bigr) \right]^{-1} \nabla \ell\bigl(\mathbf{c}^{(t)}\bigr),
	\label{eq:cut-newton}
\end{equation}
\noindent where $\mathbf{c} = (c_1, c_2, \dots, c_{K-1})'$ is the vector of cuts, $\nabla \ell(\mathbf{c})$ is the vector of first derivatives, and $\nabla^2 \ell(\mathbf{c})$ is the matrix of second derivatives. 

We iterate until no cut moves by more than a small tolerance, set by default to $10^{-3}\sigma$. In practice iteration converges in two or three steps.\footnote{As a failsafe, if an iteration lowers the conditional likelihood or puts the cuts out of increasing order, we halve the step size and try again. But this failsafe was not triggered in any of our simulations or analyses.} Convergence is fast because the first and second derivatives can be written in closed form. The first derivative vector $\nabla \ell(\mathbf{c})$ contains the $K-1$ first derivatives with respect to each cut $k$:
\begin{equation}
	\frac{\partial \ell}{\partial c_k}
	\;=\; \sum_{g=1}^G \frac{\varphi(z_{gk})}{\sigma_g}
	\left( \frac{n_{gk}}{P_{gk}} - \frac{n_{g,k+1}}{P_{g,k+1}} \right)
	\label{eq:cut-score}
\end{equation}

\noindent The second derivative matrix $\nabla^2 \ell(\mathbf{c})$ has a diagonal containing the second derivatives with respect to each cut $k$:
\begin{equation}
	\frac{\partial^2 \ell}{\partial c_k^2}
	= \sum_{g=1}^G \left[
	-\frac{z_{gk}\varphi(z_{gk})}{\sigma_g^{2}}
	\left( \frac{n_{gk}}{P_{gk}} - \frac{n_{g,k+1}}{P_{g,k+1}} \right)
	-\frac{\varphi(z_{gk})^{2}}{\sigma_g^{2}}
	\left( \frac{n_{gk}}{P_{gk}^{2}} + \frac{n_{g,k+1}}{P_{g,k+1}^{2}} \right)
	\right]
	\label{eq:cut-hess-diag}
\end{equation}
	
\noindent Just off the diagonal it contains the second derivatives with respect to cuts $k$ and $k+1$:
\begin{equation}
	\frac{\partial^2 \ell}{\partial c_k \partial c_{k+1}}
	= \sum_{g=1}^G \frac{n_{g,k+1}\,\varphi(z_{gk})\,\varphi(z_{g,k+1})}{\sigma_g^{2}\,P_{g,k+1}^{2}}.
	\label{eq:cut-hess-off}
\end{equation}

\noindent And further off the diagonals the second derivatives are zero because they involve nonadjacent cuts whose likelihoods are affected by different bins---i.e., $\partial^2 \ell / \partial c_k \partial c_j = 0$ if $|k-j| \geq 2$.

\paragraph{Starting values.}Cuts will converge to the same ML estimates from many plausible starting values, but convergence is faster if we choose starting values that are not too far from the final cuts. To choose starting values, we temporarily assume that the pooled distribution is normal and place each cut at the corresponding quantile:
\begin{equation}
	c_k^{(0)} = \mu + \sigma\,\Phi^{-1}(\hat F_k), \qquad k = 1, \dots, K-1,
	\label{eq:cutinit}
\end{equation}
\noindent where $\hat F_k = \sum_g p_g \hat F_{gk}$, with $p_g = n_g / n$, is the pooled counterpart of Equation~\eqref{eq:Fhat}---the proportion of students in the pooled distribution who scored in bins 1 through $k$. 

We should acknowledge that the distribution of pooled bin counts is not actually normal. In Texas the empirical distribution is not normal at all, but the HETOP model fits a normal distribution to each district. Pooling the fitted district distributions does not produce a normal distribution; instead, it produces a student-weighted mixture of normal distributions. However, a mixture of district normal distributions can be approximately normal if the district means are not too far apart---and the normal mixture that we fit in Texas illustrates the point (Figure~\ref{fig:tx-mixture}). So initializing the cuts to the quantiles of a pooled normal distribution can offer a good starting point and quick convergence.

%% Condensed for the page-limited (anonymous) journal version; the
%% full derivation is kept for the working paper.
\ifanon
\else
\subsection{A simpler method: Quantile matching}
A simpler method for updating the cuts is also available. Again, we initialize the cuts to the quantiles of a pooled normal distribution, but in later iterations we recognize that the pooled distribution of scores across all districts is a student-weighted mixture of district normals. Then at cut $c_k$, the CDF of the fitted normal mixture is the average of the normal district CDFs $F_{gk}$ weighted by the share of all students in each district, $p_g$:
\begin{equation}
	F_k \;=\; \sum_{g=1}^G p_g F_{gk}.
	\label{eq:mixture-cdf}
\end{equation}
\noindent Quantile matching then places each cut where this modeled proportion equals the observed proportion $\hat F_k$ of students scoring below the same cut:
\begin{equation}
	F_k \;=\; \hat F_k, \qquad k = 1, \dots, K-1 .
	\label{eq:cut-quantile}
\end{equation}

\noindent Quantile matching is not quite the same as maximizing the likelihood, but it produces very similar results. In both simulated and Texas data, quantile-based cuts came within $0.0001$ SD of ML cuts. Runtime was similar, and the estimated district means and SDs were practically identical.

\fi

\section{Testing within-group normality} \label{app:gof}

\label{sec:methods-gof}

The HETOP model assumes that scores are normally distributed within each district. This assumption may be wrong, and it can be tested if there are at least 4 bins.

After estimating $(\hat\mu_g, \hat\sigma_g)$, we form the Pearson chi-square statistic:
\begin{equation}
	X^2_g \;=\; \sum_{k=1}^K \frac{\bigl(n_{gk} - n_g \hat P_{gk}\bigr)^2}{n_g \hat P_{gk}},
	\label{eq:gof-chisq}
\end{equation}
\noindent which compares the observed counts $n_{gk}$ to the expected counts $n_g \hat P_{gk}$ implied by the fitted normal density. Here $\hat P_{gk} = P_{gk}(\hat\mu_g, \hat\sigma_g)$ represents the fitted normal probabilities.

Under the null hypothesis of within-group normality, $X^2_g$ has a chi-square distribution with $K - 3$ degrees of freedom. The degrees of freedom are equal to the number of bins $K$ minus 1 because the bin proportions must add to 1, minus 2 for the two estimated parameters $(\hat\mu_g, \hat\sigma_g)$ \citep{fisher1924}. 

With 4 bins, there is only 1 degree of freedom and the power of the test is limited. In a small district, failing to reject the null hypothesis does not necessarily mean that scores are normally distributed. It could mean that departures from normality became practically undetectable after the scores were binned.

\section{Relative efficiency and the fraction of missing information} \label{app:releff}

Standard errors from binned data are larger than they would be if the data were not binned. Had we observed the $n_g$ individual test scores, the SEs would have been
\begin{align}
	\mathrm{SE}_{\text{unbinned}}(\hat\mu_g)        &= \frac{\sigma_g}{\sqrt{n_g}},
	\label{eq:app-se-unbinned} \\[4pt]
	\mathrm{SE}_{\text{unbinned}}(\hat\sigma_g)     &= \frac{\sigma_g}{\sqrt{2 n_g}},
	\label{eq:app-se-unbinned-sigma} \\[4pt]
	\mathrm{SE}_{\text{unbinned}}(\log\hat\sigma_g) &= \frac{1}{\sqrt{2 n_g}} .
	\label{eq:app-se-unbinned-logsigma}
\end{align}
\noindent Then each binned SE is the corresponding unbinned SE multiplied by an \emph{inflation factor} $1/\sqrt{R}$:
\begin{align}
	\mathrm{SE}(\hat\mu_g)        &= \mathrm{SE}_{\text{unbinned}}(\hat\mu_g)\,\frac{1}{\sqrt{R_{g\mu}}},
	\label{eq:app-inflation-mu} \\[4pt]
	\mathrm{SE}(\hat\sigma_g)     &= \mathrm{SE}_{\text{unbinned}}(\hat\sigma_g)\,\frac{1}{\sqrt{R_{g\sigma}}},
	\label{eq:app-inflation-sigma} \\[4pt]
	\mathrm{SE}(\log\hat\sigma_g) &= \mathrm{SE}_{\text{unbinned}}(\log\hat\sigma_g)\,\frac{1}{\sqrt{R_{g\sigma}}}.
	\label{eq:app-inflation-logsigma}
\end{align}

\noindent Here $R$ represents the \emph{relative efficiency} of each binned estimator---i.e., the efficiency of the binned estimate relative to the efficiency that could be achieved if the data were not binned \citep{gjeddebaek1956,stevens1948,kulldorff1963}:
\begin{align}
	R_{g\mu}    &= \frac{D}{S_{\sigma\sigma}},
	\label{eq:app-R-mu} \\[4pt]
	R_{g\sigma} &= \frac{D}{2 S_{\mu\mu}} .
	\label{eq:app-R-sigma}
\end{align}

\noindent The complement of relative efficiency is the \emph{fraction of missing information}---i.e., the fraction of information lost to binning \citep{fisher1922}:
\begin{equation}
	\mathrm{FMI} \;=\; 1 - R .
	\label{eq:app-fmi}
\end{equation}

\section{Standard errors for samples versus populations} \label{app:ml-pop}

Maximum likelihood standard errors assume that we have a sample from a larger population. But in education, it is often the case that the whole population has been tested. With population data, there is no sampling error. The only uncertainty is due to binning, and it is appropriate to report a smaller SE that reflects binning only. 

If we square the standard errors, we can isolate the binning $\mathrm{SE}^2$ by subtracting the unbinned $\mathrm{SE}^2$ from the total $\mathrm{SE}^2$,
\begin{align}
	\mathrm{SE}^2_{\text{binning}}(\hat\mu_g)
	  &= \mathrm{SE}^2_{\text{total}}(\hat\mu_g) - \mathrm{SE}^2_{\text{unbinned}}(\hat\mu_g),
	\label{eq:app-se-binning-mu} \\[4pt]
	\mathrm{SE}^2_{\text{binning}}(\hat\sigma_g)
	  &= \mathrm{SE}^2_{\text{total}}(\hat\sigma_g) - \mathrm{SE}^2_{\text{unbinned}}(\hat\sigma_g),
	\label{eq:app-se-binning-sigma} \\[4pt]
	\mathrm{SE}^2_{\text{binning}}(\log\hat\sigma_g)
	  &= \mathrm{SE}^2_{\text{total}}(\log\hat\sigma_g) - \mathrm{SE}^2_{\text{unbinned}}(\log\hat\sigma_g) .
	\label{eq:app-se-binning-logsigma}
\end{align}
\noindent Or, equivalently, we can get the binning $\mathrm{SE}^2$ by multiplying the total $\mathrm{SE}^2$ by the corresponding fraction of missing information:
\begin{align}
	\mathrm{SE}^2_{\text{binning}}(\hat\mu_g)
	&= \mathrm{SE}^2_{\text{total}}(\hat\mu_g)\,\mathrm{FMI}_{g\mu},
	\label{eq:app-fmi-mu} \\[4pt]
	\mathrm{SE}^2_{\text{binning}}(\hat\sigma_g)
	&= \mathrm{SE}^2_{\text{total}}(\hat\sigma_g)\,\mathrm{FMI}_{g\sigma},
	\label{eq:app-fmi-sigma} \\[4pt]
	\mathrm{SE}^2_{\text{binning}}(\log\hat\sigma_g)
	&= \mathrm{SE}^2_{\text{total}}(\log\hat\sigma_g)\,\mathrm{FMI}_{g\sigma} .
	\label{eq:app-fmi-logsigma}
\end{align}

\section{Empirical Bayes shrinkage} \label{app:eb}

\label{sec:methods-eb}

Maximum likelihood estimates can be biased and unreliable in low-information districts where scores are concentrated in the top or bottom bin. Empirical Bayes estimates can improve unreliable estimates by shrinking them toward the grand mean.

To get an empirical Bayes estimate $\tilde\mu_g$ of the district mean, we shrink the ML estimate $\hat\mu_g$ toward the grand mean $\bar\mu$ \citep{fay1979}:
\begin{equation}
\tilde\mu_g \;=\; w_g\, \hat\mu_g \;+\; (1 - w_g)\, \bar\mu,
\label{eq:eb-mu}
\end{equation}

\noindent where the shrinkage weight $w_g$ is the reliability of $\hat\mu_g$, estimated by
\begin{equation}
	w_g \;=\; \frac{\hat\tau^2}{\hat\tau^2 + \mathrm{SE}^2(\hat\mu_g)},
	\label{eq:eb-mu-pop}
\end{equation}
\noindent 
Here $\bar\mu$ is estimated by weighting each district by $p_g$, the district's share of students tested:
\begin{equation}
	\bar\mu \;=\; \sum_g p_g\, \hat\mu_g .
	\label{eq:eb-mubar}
\end{equation}

\noindent $\mathrm{SE}(\hat\mu_g)$ is the standard error of the district mean. It may be either a sample-based standard error or the population standard error defined in Appendix~\ref{app:ml-pop}. And $\tau^2$ is the mean squared deviation of the true district means from $\bar\mu$. We estimate $\tau^2$ by method of moments, subtracting the mean squared standard error from the mean squared deviation of the ML estimates:
\begin{equation}
	\hat\tau^2 \;=\; \frac{1}{G}\sum_g \left(\hat\mu_g - \bar\mu\right)^2 \;-\; \frac{1}{G}\sum_g \mathrm{SE}^2(\hat\mu_g).
	\label{eq:eb-tau2}
\end{equation}

\noindent On rare occasions, this formula can produce a negative estimate, which we round up to zero.

EB also shrinks the standard error, but only by the square root of the weight applied to the point estimate \citep{morris1983},
\begin{equation}
	\mathrm{SE}(\tilde\mu_g) \;=\; \sqrt{w_g}\;\mathrm{SE}(\hat\mu_g),
	\label{eq:eb-se-mu}
\end{equation}
\noindent and 95\% confidence intervals add and subtract 2 shrunken standard errors from the shrunken point estimate. 

The SD is shrunk the same way on the log scale, substituting $\log \hat\sigma_g$ for $\hat\mu_g$ throughout.

\end{document}